\documentclass[a4paper,fleqn,usenatbib]{mnras}
\usepackage{newtxtext,newtxmath}
\usepackage[T1]{fontenc}
\usepackage{ae,aecompl}
\usepackage{graphicx}	
\usepackage{amsmath}	
\usepackage{amssymb}	
\usepackage{longtable}
\usepackage{deluxetable}
\usepackage{enumitem}

\newcommand{\oiii}{[O\,{\sc iii}]}
\newcommand{\ovi}{O\,{\sc vi}}

\newcommand{\feii}{Fe\,{\sc ii}}

\newcommand{\siv}{S\,{\sc iv}}
\newcommand{\siiv}{Si\,{\sc iv}}

\newcommand{\civ}{C\,{\sc iv}}
\newcommand{\mgii}{Mg\,{\sc ii}}

\newcommand{\angstrom}{\text{ \normalfont\AA}}

\def\lya{Ly$\alpha$}
\def\ly{$\lambda$}

\def\hb{H$\beta$}

\def\civ{C\,{\sc iv}}
\def\nii{N\,{\sc ii}}

\def\nv{N\,{\sc v}}

\def\oiii{O\,{\sc iii}}

\def\ovi{O\,{\sc vi}}

\def\mgii{Mg\,{\sc ii}}

\def\siiv{Si\,{\sc iv}}

\def\siv{S\,{\sc iv}}

\def\feii{Fe\,{\sc ii}}

\def\Ek{$\dot{E}_{k}$}

\def\nh{\ifmmode n_\mathrm{\scriptscriptstyle H} \else $n_\mathrm{\scriptscriptstyle H}$\fi}
\def\ne{\ifmmode n_\mathrm{\scriptstyle e} \else $n_\mathrm{\scriptstyle e}$\fi}
\def\Qh{\ifmmode Q_\mathrm{\scriptstyle H} \else $Q_\mathrm{\scriptstyle H}$\fi}
\def\Uh{\ifmmode U_\mathrm{\scriptstyle H} \else $U_\mathrm{\scriptstyle H}$\fi}
\def\Nh{\ifmmode N_\mathrm{\scriptstyle H} \else $N_\mathrm{\scriptstyle H}$\fi}
\def\Uhhp{\ifmmode U_\mathrm{\scriptstyle H,HP} \else $U_\mathrm{\scriptstyle H,HP}$\fi}
\def\Nhhp{\ifmmode N_\mathrm{\scriptstyle H,HP} \else $N_\mathrm{\scriptstyle H,HP}$\fi}
\def\Uhvhp{\ifmmode U_\mathrm{\scriptstyle H,VHP} \else $U_\mathrm{\scriptstyle H,VHP}$\fi}
\def\Nhvhp{\ifmmode N_\mathrm{\scriptstyle H,VHP} \else $N_\mathrm{\scriptstyle H,VHP}$\fi}
\def\Nion{\ifmmode N_\mathrm{\scriptstyle ion} \else $N_\mathrm{\scriptstyle ion}$\fi}

\def\Zsun{\ifmmode {\rm Z}_{\odot} \else Z$_{\odot}$\fi}
\def\Msun{\ifmmode {\rm M}_{\odot} \else M$_{\odot}$\fi}
\def\kms{\ifmmode {\rm km~s}^{-1} \else km~s$^{-1}$\fi}
\def\Lya{\ifmmode {\rm Ly}\alpha \else Ly$\alpha$\fi}
\def\Lyb{\ifmmode {\rm Ly}\beta \else Ly$\beta$\fi}
\def\Lyg{\ifmmode {\rm Ly}\gamma \else Ly$\gamma$\fi}
\def\Lyd{\ifmmode {\rm Ly}\delta \else Ly$\delta$\fi}
\def\neaod{\ifmmode n_\mathrm{\scriptscriptstyle AOD} \else $n_\mathrm{\scriptscriptstyle AOD}$\fi}
\def\necrit{\ifmmode n_\mathrm{\scriptstyle cr} \else $n_\mathrm{\scriptstyle cr}$\fi}
\def\ncr{\ifmmode n_\mathrm{\scriptstyle cr} \else $n_\mathrm{\scriptstyle cr}$\fi}
\def\nepi{\ifmmode n_\mathrm{\scriptscriptstyle PI} \else $n_\mathrm{\scriptscriptstyle PI}$\fi}
\def\gtorder{\mathrel{\raise.3ex\hbox{$>$}\mkern-14mu\lower0.6ex\hbox{$\sim$}}}
\def\ltorder{\mathrel{\raise.3ex\hbox{$<$}\mkern-14mu\lower0.6ex\hbox{$\sim$}}}
\def\vmed{$v_{\text{med}}$}
\def\vro{\ifmmode v_\mathrm{\scriptscriptstyle 1, \scriptstyle r} \else $v_\mathrm{\scriptscriptstyle 1, \scriptstyle r}$\fi}
\def\vrc{\ifmmode v_\mathrm{\scriptscriptstyle 2, \scriptstyle r} \else $v_\mathrm{\scriptscriptstyle 2, \scriptstyle r}$\fi}
\def\vzo{\ifmmode v_\mathrm{\scriptscriptstyle 1, \scriptstyle z} \else $v_\mathrm{\scriptscriptstyle 1, \scriptstyle z}$\fi}
\def\vzc{\ifmmode v_\mathrm{\scriptscriptstyle 2, \scriptstyle z} \else $v_\mathrm{\scriptscriptstyle 2, \scriptstyle z}$\fi}

\newcommand{\WOIII}{W$_{\text{[OIII]}}$}
\newcommand{\WCIV}{W$_{\text{CIV}}$}

\title[Evidence that Emission and Absorption Outflows in Quasars Are Related]{Evidence that Emission and Absorption Outflows in Quasars Are Related}

\author[Xu et al.]{
Xinfeng Xu,$^{1}$\thanks{E-mail: xinfeng@vt.edu}
Nadia L. Zakamska$^{2}$,
Nahum Arav$^{1}$,
Timothy Miller$^{1}$,
Chris Benn$^{3}$
\\
$^{1}$Department of Physics, Virginia Tech, Blacksburg, VA 24061, USA\\
$^{2}$Department of Physics \& Astronomy, Johns Hopkins University, Bloomberg Center, 3400 N. Charles St., Baltimore, MD 21218, USA\\
$^{3}$Isaac Newton Group of Telescopes, La Palma, Canary Islands, Spain
}

\date{Submitted Dec. 18th, 2019}

\pubyear{2019}

\hypersetup{draft} 
\begin{document}
\label{firstpage}
\pagerange{\pageref{firstpage}--\pageref{lastpage}}

\maketitle	

\begin{abstract}
We analyze VLT/X-shooter data for 7 quasars, where we study the relationships between their broad absorption line (BAL) and emission line outflows. We find: 1) the luminosity of the [\oiii] \ly 5007 emission profile decreases with increasing electron number density (\ne) derived from the BAL outflow in the same quasar, 2) the measured velocity widths from the [\oiii] emission features and \civ\ absorption troughs in the same object are similar, and 3) the mean radial velocity derived from the BAL outflow is moderately larger than the one from the [\oiii] emission outflow. These findings can be explained by the physical interpretation that the [\oiii] and BAL outflow are different manifestations of the same wind. When we have outflows with smaller distances to the central source, their \ne\ is higher. Therefore, the [\oiii] emission is collisionally de-excited and the [\oiii] luminosity is then suppressed. Comparisons to previous studies show that the objects in our sample exhibit broad [\oiii] emission features similar to the ones in extremely red quasars (ERQs). This might imply that BAL quasars and ERQs have the same geometry of outflows or are at a similar evolutionary stage. We found that the physical parameters derived from the BAL outflows can explain the amount of observed [\oiii] luminosity, which strengthens our claim of both BAL and [\oiii] outflows are from the same wind. These estimates can be tested with upcoming James Webb Space Telescope observations.


\end{abstract}




\section{Introduction}
\label{sec:intro}
``Quasar-mode feedback'' occurs when momentum and energy from the environment of accreting supermassive black hole (SMBH) couple to the host galaxy \citep[e.g.,][]{Silk98,Ostriker10}. One mechanism for such a coupling is by high-velocity (up to $\sim$ 0.2c) quasar-driven ionized outflows, which propagate into the interstellar medium of the host galaxy. Given enough energy and momentum, these outflows are capable of affect their host galaxies. Such quasar outflows are invoked to explain a variety of observations, e.g, the chemical enrichment of the intergalactic medium (IGM) \citep[e.g.,][]{Scannapieco04, Moll07, Khalatyan08, Baskin12, tay15,tay17}, the shape of the observed quasar luminosity function \citep[e.g.,][]{Hopkins05, Hopkins07, Hopkins10, Han12, Singal13,Wada15}, and the self-regulation of the growth of the SMBHs \citep[e.g.,][]{Silk98, Hopkins16, Ciotti17, Angles17,Beckmann17}.
Recent cosmological hydrodynamical simulations confirm that quasar outflows can remove and prevent new accretion of cold gas, and therefore, quench star formation and transform the hosts from blue compact galaxies to red extended galaxies \citep{Hopkins16, Choi18}. 

Considerable observational effort has been directed through multiple observational techniques for studying different phases of quasar outflows, including molecular outflows \citep[e.g.,][]{Nesvadba10, Rupke13, Sun14,Veilleux17,Brusa18,Bischetti19}, [\oiii] emission outflows \citep[e.g.,][]{Alexander10, Liu13a, Liu13b, Zakamska16,Perrotta19,Temple19}, ultraviolet (UV) absorption outflows \citep[e.g.,][]{Hall02,Arav13,Arav18, Grier15,Leighly18,Hamann19, Xu18, Xu19a}, and X-ray emission outflows \citep[e.g.,][]{Ogle10, Greene14,Lansbury18,Bianchi19}. It is currently not known how the different phases of the outflow are related to each other and which one carries most of the mass, energy or momentum.

Broad absorption lines (BALs) are observed in $\simeq$ 20\% of the total quasar population \citep[][]{ Hall02,Tolea02, Reichard03, Trump06, Ganguly08,Gibson09} and are indicative of ionized quasar outflows \citep[e.g.,][]{Arav99, Arav01, Hall02, Arav08, Grier15, Leighly18, Hamann19}. In the rest-frame UV, quasar absorption outflows show troughs from different ionic transitions, e.g., \lya\ \ly 1215.67, \nv\ \ly\ly 1238.82, 1242.80, \siiv\ \ly\ly 1393.75, 1402.77, and \civ\ \ly\ly 1548.19, 1550.77. To quantify the extent that absorption outflows can contribute to quasar feedback, we need to determine their kinetic luminosity (\Ek). Theoretical models predict that the ratio of \Ek\ to the Eddington luminosity of the quasar, i.e., $\Gamma_{\text{Edd}}$ $\equiv$ \Ek/L$_{\text{Edd}}$, of at least 0.5\% \citep{Hopkins10} or 5\% \citep[][]{Scannapieco04} is required for strong quasar-mode feedback effects to the host galaxies. \Ek\ and $\Gamma_{\text{Edd}}$ are proportional to the distance ($R$) of the outflow to the central source. 

There is no consensus about the $R$ values in BAL outflows and the resulting estimates of $\Gamma_{\text{Edd}}$. Accretion disk wind models predict that some BAL features likely arise near the nucleus \citep[$\sim$ 0.01 pc, e.g.,][]{Murray95, Proga00, Proga04, Higginbottom14}, which leads to small \Ek\ and $\Gamma_{\text{Edd}}$. Photoionization analysis of BAL outflows demonstrates that BAL outflows are situated at hundreds \citep[e.g.,][]{Chamberlain15a, Xu18} and even thousands of parsecs \citep[e.g.,][]{Borguet13} from the source, and that half of the outflows are at $R$ $>$ 100 pc \citep{Arav18,Xu19a}. Outflows with significant \Ek\ and $\Gamma_{\text{Edd}}$ were also found, i.e., \Ek\ up to 5 $\times$ 10$^{46}$ erg s$^{-1}$ and $\Gamma_{\text{Edd}}$ up to 20\% \citep{Xu19b}. These BAL outflows with large $R$ were suggested to be formed ``\textit{in situ}'' by radiative shocks of the interstellar cloud by a quasar hot wind \citep[e.g.,][]{Faucher12}. Our group's recent hydro-dynamical simulations also show that this ``\textit{in situ}'' absorbing clouds can reproduce the observed BAL quasar outflows properties \citep[e.g.,][]{Meir19}.


Another commonly used tracer of ionized quasar outflows is forbidden-line emission features, e.g., from [\nii] \ly\ly 6549.86, 6585.27, and [\oiii] \ly\ly 4958.91, 5006.84. Since these features are suppressed by collisions at high density, low gas densities (n $\lesssim$ 7 $\times$ 10$^{5}$ cm$^{-3}$) are needed for generating forbidden lines \citep{Baskin05}. Integral Field Unit (IFU) observations \citep[e.g.,][]{Liu13a, Liu13b} show that forbidden-line emission features arise from distances of 0 -- $\sim$ 30 kpc from the quasar. But the structure of outflows on $<$ 3 kpc scales is not well-known since it is the resolution limit of ground-based observations \citep{Liu13a}.



\begin{table}
	\centering
	\caption{VLT/X-Shooter Observation Details}
	\label{tab:example_table}
	\begin{tabular}{lccccr} 
		\hline
		\hline
		Object$^{a}$ (SDSS) 	& 	Ref.$^{b}$	&	z$^{c}$ 	& 	r$^{d}$  &	BAL$^{e}$	& 	Exp.$^{f}$\\
		\hline
		J0046+0104  		& 	1	&2.149 & 18.04  &	B	& 10.8\\
		J0825+0740  		& 	1	&2.204 & 17.89  &	M	& 18.0\\
		J0831+0354 		& 	2	&2.076 & 18.27  &	B	& 10.8\\
		J0941+1331  		& 	1	&2.021 & 18.15  &	B	& 10.8\\
		J1111+1437 		& 	3	&2.132 & 18.03  &	M	& 10.8\\
		J1135+1615  		& 	1	&2.004 & 18.36  &	B	& 7.2\\
		J1512+1119  		& 	4	&2.109 & 17.65  &	B	& 8.4\\


		\hline
	\multicolumn{5}{l}{%
  	\begin{minipage}{7cm}%
	Note. --\\
    	(a)\ \ Objects right ascension and declination.\\
	(b) Analysis of absorption outflows for these objects are reported in: 1: \cite{Xu19a}; 2: \cite{Chamberlain15b}; 3: \cite{Xu18}; 4: \cite{Borguet12b}.\\
	(c) Redshifts for the quasars obtained from matching the \mgii\ \ly 2800\ emission lines \citep{Hewett10,Shen16}.\\ 
	(d) $r$-band magnitude obtained through PSF fitting.\\
	(e) B for BAL and M for mini-BAL.\\ 
	(f) Total exposure time in kilo-seconds.\\
  	\end{minipage}%
	}\\
	\end{tabular}
	\\ [0mm]
	
\end{table}

Various studies show that a strong asymmetry in forbidden-line emission features, especially blue-shifts, is evidence of high-velocity extended outflows \citep[e.g.,][]{Zakamska14, Brusa15a, Perrotta19, Perna2019}. This kinematic evidence for outflows is seen in [Ne III] and [Ne V] in ultraluminous infrared galaxies \citep[][]{Spoon09}, in [\oiii] from optically selected AGNs and X-ray obscured red quasars \citep[][respectively]{Mullaney13,Brusa15a}, and in [\oiii] from type 2 quasars or Extremely Red Quasars (ERQs) \citep{Zakamska14, Zakamska16, Perrotta19}.


Since both BAL features and the forbidden-lines trace ionized gas, a direct comparison between these two diagnostics is important for understanding the relationships between different phases of the outflow. Studies of both absorption and forbidden-line emission features over a quasar sample do not exist in the literature. This is mainly due to the difficulty of obtaining high-quality data covering both the rest-frame regions of 1000 -- 2000\angstrom\ and 4000 -- 7000\angstrom\ in the same object. Analyses of absorption and forbidden-line emission features in a single object are also rare [only one IFU-related study in \cite{Liu15} and spectra-related study in \cite{Tian19}]. They show that the distances and kinematics revealed by the [\oiii] emission features are roughly consistent with UV absorption-line analyses in the same quasar.

In this paper, we analyze high-quality observations of quasar outflows seen in both absorption and forbidden-line emission features in the same object. The sample comes from Very Large Telescope (VLT)/X-Shooter observations (see description of X-Shooter in section \ref{sec:obs}) of 7 BAL/mini-BAL quasars at redshift z $\sim$ 2, for which the absorption outflow's distance ($R$) from the central source is measured \citep[][]{Borguet12b,Chamberlain15b,Xu18,Xu19a}. As described in section \ref{sec:obs}, due to the wide spectral coverage (3000 -- 25,000\angstrom) of VLT/X-Shooter,  we also observed forbidden-line emission features from [\oiii] \ly\ly 4959, 5007. Here we focus on modeling and analyzing the observed [\oiii] emission features in these quasars, as well as exploring the relationships between absorption outflows and forbidden-line emission outflows. 



The structure of the paper is as follows. In Section \ref{sec:obs}, we present the observations and data reduction. In Section \ref{sec:analysis}, we use multiple methods to analyze the kinematics of forbidden-line emission features from [\oiii]. We show the relationships between the observed absorption outflows and the forbidden-line emission outflows in Section \ref{sec:connect}. In Section \ref{sec:discuss}, we propose possible geometry for observing both BAL and emission outflows in the same quasar. We also compare our results to previous studies and estimate the [\oiii] emission region's size and global covering factor in Section \ref{sec:discuss}. We summarize the paper in section \ref{sec:summary}. 

We adopt a cosmology with H$_{0}$ = 69.6 km s$^{-1}$ Mpc$^{-1}$, $\Omega_m$ = 0.286, and $\Omega_{\Lambda}$ = 0.714, and we use Ned Wright's Javascript Cosmology Calculator website \citep{Wright06}. We use air wavelengths for emission line identifications in the text, e.g., [\oiii] \ly\ly 4958.91, 5006.84 (hereafter, [\oiii] \ly 4959 and [\oiii] \ly 5007). However, since the spectra from VLT are calibrated for vacuum wavelengths, we use vacuum wavelengths for emission lines in all calculations, e.g., [\oiii] \ly\ly 4960.29, 5008.24. 

\begin{figure*}
	\includegraphics[width=2.0\columnwidth,trim={2cm 1cm 2cm 0.5cm}]{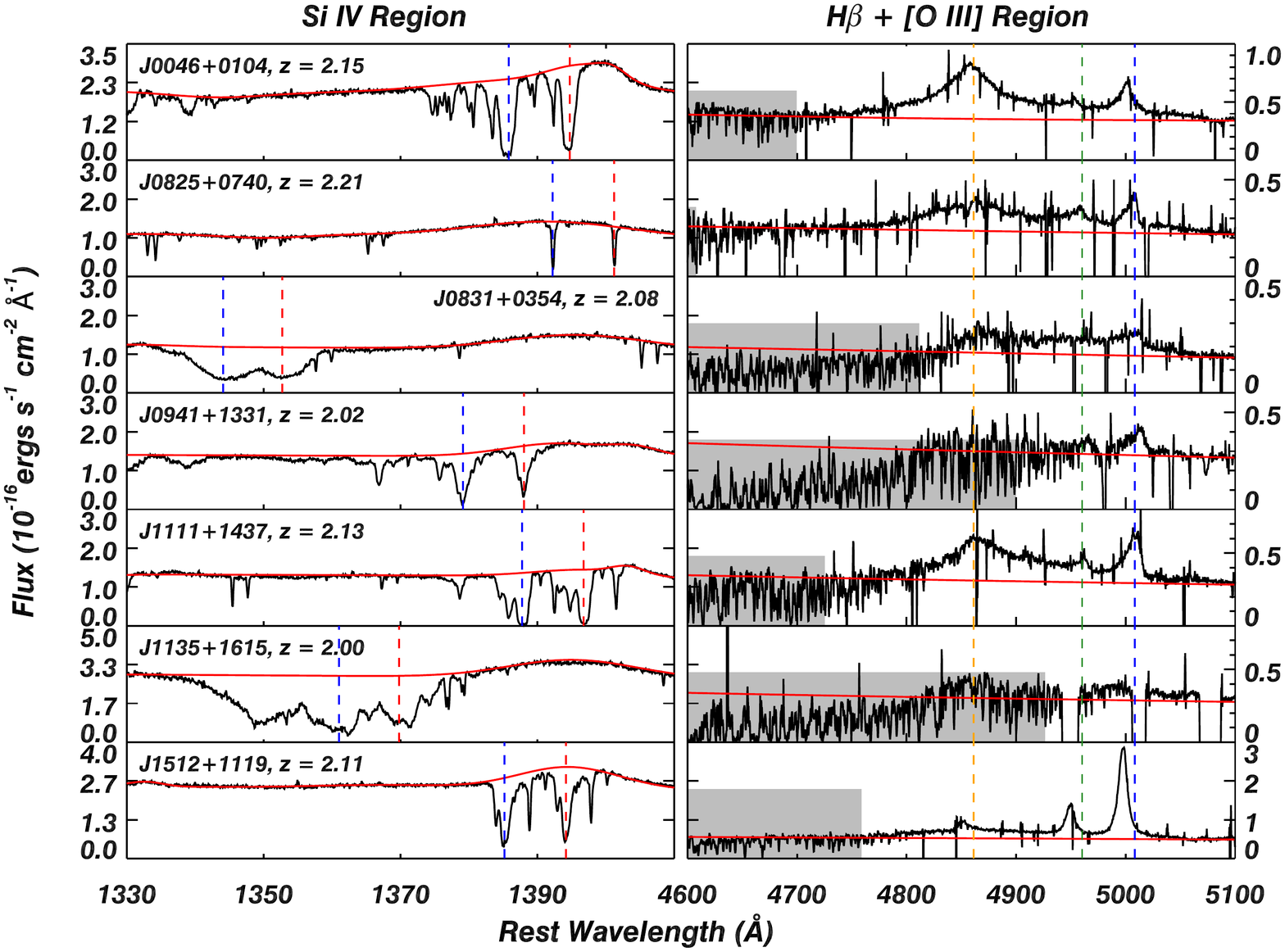}
    \caption{Comparisons between the observed \siiv\ absorption outflow regions (\textbf{Left}) and the \hb\ + [\oiii] forbidden-line emission outflow regions (\textbf{Right}). Each row shows a single object. All 7 objects have significant \siiv\ absorption outflows, and at least 6 of them have significant [\oiii] emission outflows. We mark the regions which are contaminated by atmospheric water absorption features in gray shades. SDSS J1135+1615 has instrumental artifacts near 4950\angstrom\ and 5010\angstrom\ rest-frame region. The blue and red dashed lines in the left panel show the absorption troughs' velocity centroids for the \siiv\ absorption outflows, for which $R$ is determined (see section \ref{sec:PPBAL}). The orange, green, and blue dashed lines in the right panel show the expected location of \hb, [\oiii] \ly 4960.29, and \ly 5008.24 in the quasar's rest-frame, respectively.}
    \label{fig:OIII_all}
\end{figure*}

\begin{figure*}
	\includegraphics[width=1.00\columnwidth,angle=0,clip=true,trim={0.0cm 1.8cm 0cm 0cm}]{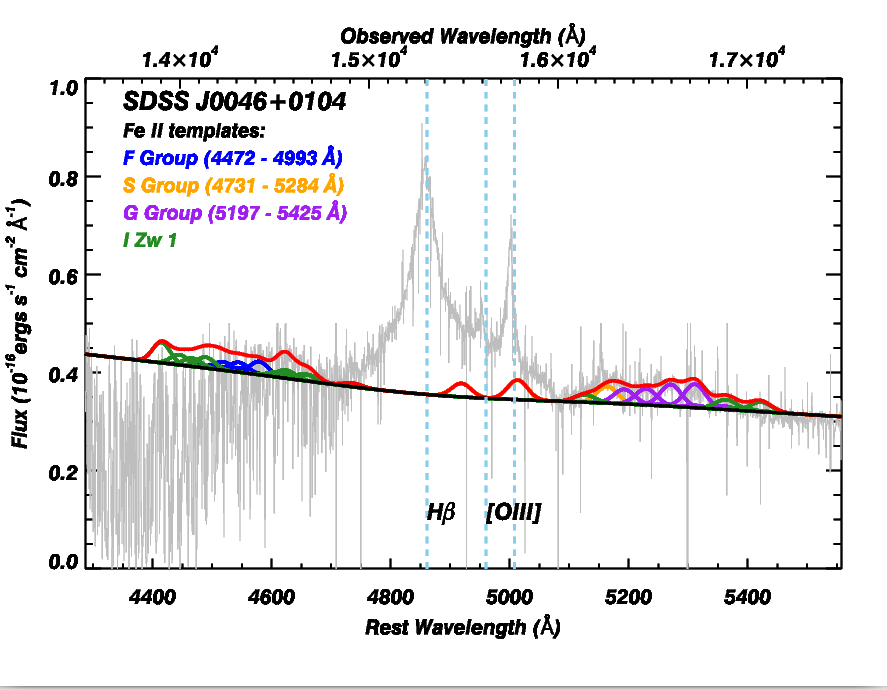}
	\includegraphics[width=1.00\columnwidth,angle=0,clip, trim={0.0cm 0.9cm 1.8cm 0.5cm}]{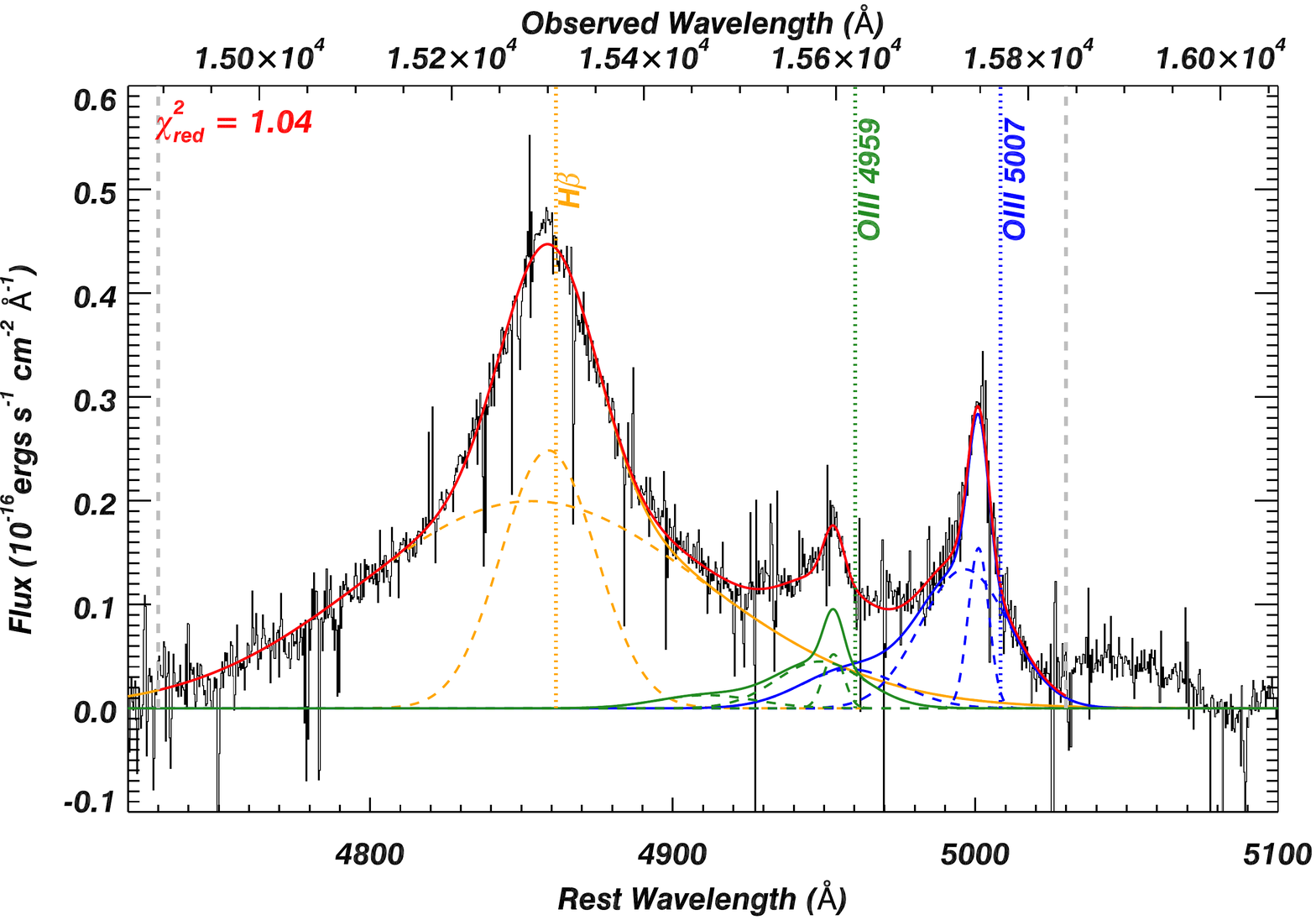}\\
	\noindent\hspace{0cm}\large\textbf{(a)}\noindent\hspace{8cm}\textbf{(b)}\\
	\includegraphics[width=1.0\columnwidth,angle=0,clip,	trim={0.0cm 0.9cm 1.8cm 2.0cm}]{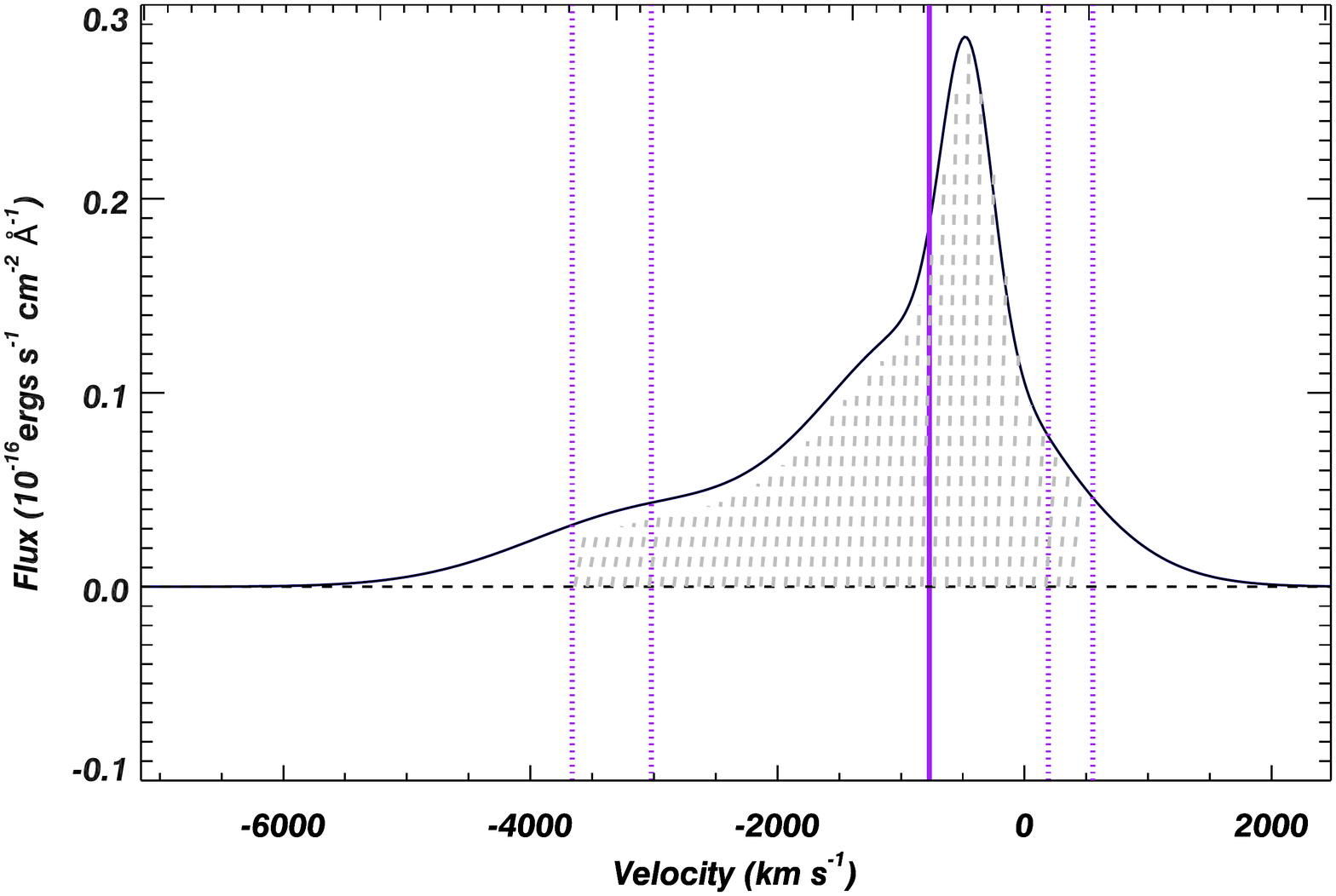}
	\includegraphics[width=1.0\columnwidth,angle=0,clip,	trim={0.6cm 0.0cm 2.0cm 1.0cm}]{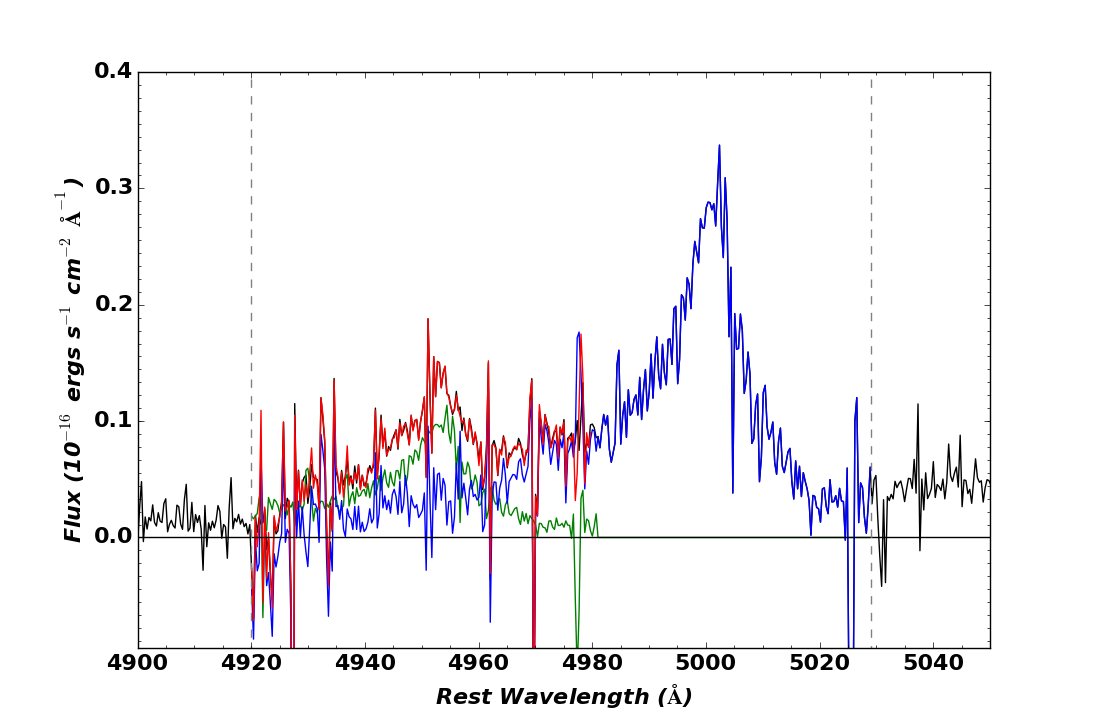}\\
	\noindent\hspace{0cm}\large\textbf{(c)}\noindent\hspace{8cm}\textbf{(d)}
	\caption[caption]{[\oiii] analyses for quasar SDSS J0046+0104.
 
\textbf{Panel (a):} The best fitting \feii\ template. \feii\ emission profiles are divided into four groups, and are plotted in different colors (see section \ref{sec:FeII} for details). The continuum level is the same as the one in figure \ref{fig:OIII_all} and is shown as the black solid line. The overall \feii\ model created by summing all four groups is shown as the solid red line. The blue dotted lines (left to right) indicate the expected wavelength locations of \hb, [\oiii] \ly 4959, and [\oiii] \ly 5007 in the quasar's rest-frame, respectively. Atmospheric absorption features exist at $\lesssim$ 15,000\angstrom\ in the observed-frame. 

\textbf{Panel (b):} The fits to the \hb\ + [\oiii] emission lines region, where we have already subtracted the continuum (figure \ref{fig:OIII_all}) and the fitted \feii\ emission features (panel (a)). We use 2 or 3 Gaussian profiles to fit the \hb\ and [\oiii] emission profiles. The Gaussian models for \hb, [\oiii] \ly 4959, [\oiii] \ly 5007 are shown in orange, green, and blue dashed lines, respectively. The overall model created by summing up all Gaussian profiles is shown as the solid red line. We mark the expected wavelength locations of \hb, [\oiii] \ly 4959, and [\oiii] \ly 5007 in the quasar rest-frame in orange, green, and blue dotted lines, respectively. The fitting range is within the two gray dashed lines. The reduced $\chi^{2}$ value of the fitting to the data is shown at the top-left corner. See detailed descriptions in section \ref{sec:OIII}. 

\textbf{Panel (c):} The synthetic profile of the [\oiii] \ly 5007 emission (in black, obtained in section \ref{sec:OIII}) relative to quasar's rest-frame velocity. The dotted purple lines represent the v$_{05}$, v$_{10}$, v$_{90}$, and v$_{95}$ (from left to right), while the solid purple line is the median velocity of the profile, i.e., v$_{50}$. The gray slanted region marks the part of the profile containing 90\% of the [\oiii] \ly 5007 emission line power, i.e., $w_{90}$. 

\textbf{Panel (d):} The non-parametric fits to the [\oiii] emission region (black line), where we have already subtracted the continuum (figure \ref{fig:OIII_all}), the fitted \feii\ emission feature (panel (a)), and the modeled \hb\ emission feature (pabel (b)). We show the best non-parametric fits where the green and blue curves are the models for the [\oiii] \ly 4959 and [\oiii] \ly 5007, respectively. These models are consistent with the parametric models in Panel (b). The overall model by summing [\oiii] \ly 4959 and [\oiii] \ly 5007 emission are plotted in red between 4920 -- 4980\angstrom\ rest-frame. We mark the fitting range with dashed gray lines. See section \ref{sec:nonpara} for discussion.}
    \label{fig:J0046all}
\end{figure*}

\section{Observations and Data Reduction}
\label{sec:obs}
We carried out several VLT/X-shooter programs (PI: Benn) for quasars with z between $\sim$ 2 -- 2.6. The details of the observations that we analyze here are shown in table \ref{tab:example_table}. These objects are selected from Sloan Digital Sky Survey (SDSS)/BOSS by two criteria \citep[][]{Xu19a}: 1) r band magnitude $\lesssim$ 18.8, and 2) deep \siiv\ \ly\ly 1393.76, 1402.77 troughs. The latter criterion ensures a higher probability of detecting the \siv\ \ly 1062.66 trough in VLT/X-Shooter spectra, where the depth ratio between \siv\ \ly 1062.66 and \siv*\ \ly 1072.97 troughs is related to $R$ (see section \ref{sec:PPBAL}). SDSS covers the wavelength range between 3800\angstrom\ and 10400\angstrom, while BOSS covers 3650\angstrom\ to 10400\angstrom. For objects with redshifts $<$ 2.6, the \siv*\ and \siv\ troughs are not detectable in SDSS/BOSS spectral range. Therefore, choosing quasars from SDSS and BOSS with z $<$ 2.6 guarantees that we would not have a bias toward specific depth ratios of the \siv*\ and \siv\ troughs, which creates a bias for certain $R$ values. Therefore, we call this survey a ``Blind Survey'', where a priori we had no information on the \siv\ absorption. We select objects with redshift $>$ 2 such that the absorption troughs from \siv\ and \siv* are covered by VLT/X-Shooter.

X-Shooter is a second-generation instrument on VLT. It is a medium resolution (R $\sim$ 6000 -- 9000) spectrograph with a wide spectral coverage (3000 -- 25,000\angstrom), which observes both the absorption-line and emission-line outflow diagnostics. Here we concentrate on objects that show \siv\ outflow troughs (7 out of 14 quasars). For these outflows, we have already determined their total column density (\Nh), electron number density (\ne), distance from the central source ($R$), and kinetic luminosity (\Ek) \citep[][and see the table 3 of \cite{Xu19a} for a summary]{Borguet12b, Chamberlain15b, Xu18}. We describe the extraction methods of these parameters in section \ref{sec:PPBAL}. 



We reduced the data following the same method described in \cite{Xu18}. The redshifts for these quasars are obtained from matching the \mgii\ \ly 2800\ emission lines \citep[e.g.,][]{Hewett10}. \cite{Shen16} showed that \mgii\ has a mean offset of only $\sim$ 100 km s$^{-1}$ from narrow [\oiii] lines. Therefore, \mgii\ \ly 2800\ is also a good redshift indicator for the [\oiii] region. In figure \ref{fig:OIII_all}, we show the comparisons of the absorption troughs from \siiv\ to the forbidden-line emission region from [\oiii] for each object. The continuum is modeled by a single power law, and we fit the \siiv\ emission lines with 2 -- 3 Gaussian profiles. For each object, the red line represents the modeled continuum + \siiv\ emission features. The \hb\ and [\oiii] emission profiles carry the emission outflow signatures and we show their analysis in section \ref{sec:analysis}. The velocity centroids of the absorption outflows are determined by the deepest point in the absorption feature of the observed \siiv \ly\ly 1393.76, 1402.77 troughs, which is associated with the \siv\ \ly 1062.66 trough \citep[see figure 2 in ][]{Xu19a}. 

As shown in table 2 of \cite{Xu19a}, based on the \civ\ absorption troughs widths, 5 out of the 7 objects are classified as BAL quasars \citep{Weymann91}. The other two objects are classified as mini-BALs \citep[e.g.,][]{Churchill1999}, but their \civ\ widths are close to the BAL definitions. For convenience, we refer this sample as ``BAL'' throughout the remainder of this paper.



\section{Spectral Analysis}
\label{sec:analysis}

To measure the kinematics of the [\oiii] emission features in our objects, we begin by quantifying the possible contaminations from \feii\ emission in section \ref{sec:FeII}. In section \ref{sec:OIII}, we show the Gaussian fitting procedure to the \hb\ + [\oiii] region and [\oiii] measurements. Finally in section \ref{sec:nonpara}, we show the non-parametric fittings to the [\oiii] emission features.


\subsection{\feii\ Emission Templates}
\label{sec:FeII}
Optical \feii\ emission lines (4400 -- 5500\angstrom) arise from numerous transitions of the complex Fe$^{+}$ ion, and can possibly contaminate the [\oiii] \ly\ly 4959, 5007 emission region. The origin of \feii\ lines, mechanisms of their excitation, and wavelengths of the transitions in quasars are still open questions \citep[e.g.,][]{Baldwin04, Bruhweiler08}. There are various studies which created \feii\ templates from different prospectives: 1) observational studies of \feii-rich quasar spectra, especially I Zw 1 \citep[PG 0050+124, e.g.,][]{Boroson92, Veron04}; 2) theoretical calculation of the emission lines by including a large number of \feii\ atomic levels, e.g., \cite{Verner99, Bruhweiler08}, who reported 371 and 830 energy levels of \feii, up to 11.6 eV and 14.06 eV, respectively. 

To quantify and remove the contribution from possible \feii\ emission lines, we follow the \feii\ template method reported in \cite{Kova10}, where they combine the observational \feii\ with the atomic data. They divided the clearly identified \feii\ emission lines into three groups based on their lower energy levels, e.g., transitions from the lower energy levels of $^{4}$F, $^{4}$S, and $^{4}$G belong to F, S, and G group, respectively. All other lines, whose origins are not well known but have been observed in I Zw 1, belong to the Z group \citep[][]{Kova10}.

We adopt the same four-group \feii\ template as follows. The template has 7 fitting parameters, and we fit the observed \feii\ emission features with Gaussian profiles. The first two parameters are the velocity centroids and widths, which are shared by all the groups. The next four parameters are the relative intensities for each transition within the F, S, G, and Z group. For transitions in Z group, their relative intensities are obtained from I Zw 1. The seventh parameter is the excitation temperature used in calculating the relative intensities \citep[][]{Kova10}.  

By assuming T = 15,000 K \citep{Kova10} and adopting the relative intensity values from their tables 1 and 2, we show examples of the best fitting \feii\ models to the \hb\ + [\oiii] region in the panel (a) of figure \ref{fig:J0046all} (the other objects have similar spectra and are shown in Appendix \ref{sec:APPA}). The data are shown in gray while the four different \feii\ groups are shown as the solid lines with different colors. The overall \feii\ models are in solid red lines. Two strong \feii\ emission lines of the S group, \feii\ \ly 4923.93 and \feii\ \ly 5018.44, are within the \hb\ + [\oiii] region from 15,000\angstrom\ to 16,000\angstrom\ (observed-frame). Their intensities are constrained by fittings other S group lines (in orange solid lines), e.g., \feii\ \ly 5169.33 at $\sim$ 16,250\angstrom\ in the observed-frame.

\subsection{[\oiii] Kinematics}
\label{sec:OIII}
\subsubsection{Gaussian Fitting to the [\oiii] Emission Features}
\label{subsec:OIII}
After quantifying the contributions from the \feii\ emission lines, we subtract them as well as the continuum flux from the spectra. Then we measure the [\oiii] kinematics adopting the Gaussian fitting procedure of \cite{Zakamska16}. There are atmospheric absorption features in the $\lesssim$ 15,000\angstrom\ observed-frame region (marked by the gray shades in figure \ref{fig:OIII_all}). They affect mainly the broad \hb\ emission features, and for two of our objects (SDSS J0941+1331 and SDSS J1135+1615), they affect the [\oiii] emission region as well. The derived physical parameters for these two objects have large error bars which account for these contaminations. 

We use 2 or 3 Gaussian profiles to fit the observed \hb, [\oiii] \ly 4959, and [\oiii] \ly 5007 emission features. A reduced chi-square ($\chi^{2}_\text{red}$) value is calculated between the data and the fitted Gaussian profiles. If adding the third Gaussian component leads to a decrease of $\chi^{2}_\text{red}$ by $>$ 10\%, we accept the fit with a 3-Gaussian profile. Otherwise, a 2-Gaussian profile is adopted. We assume that [\oiii] \ly 4959 and [\oiii] \ly 5007 have the same kinematics, i.e., the same velocity centroids and widths for their Gaussian profiles. The intensity ratio between [\oiii] \ly 4959 and [\oiii] \ly 5007 is fixed at 1:2.99, respectively \citep{Osterbrock81}. We then explore the two different fitting scenarios to the \hb\ emission features: 1) a ``kinematically tied" model, which assumes that the \hb\ has the same kinematics as the [\oiii] doublet; 2) a ``kinematically untied" model, where \hb\ can have a different kinematic structure than the [\oiii] doublet. We find that all of our objects favor the latter scenario, since in most cases, the \hb\ emission features are less blue-shifted than the [\oiii] emission features. 

We show the fitting result for quasar SDSS J0046+0104 in the panel (b) of figure \ref{fig:J0046all} (fits for the remaining objects are shown in Appendix \ref{sec:APPA}). Individual Gaussian profiles are shown as the color dashed lines, while the combinations of them for each line are shown as solid colored lines. The overall model created by summing up all Gaussian profiles is shown as the solid red line. The 1$\sigma$ uncertainties of the parameters are computed from the covariance matrix in the fitting process. The fitted parameters with their corresponding errors are shown in part (b) of table \ref{tab:compare}. All objects except SDSS J1512+1119 show wide and blue-shifted asymmetric [\oiii] emission features. 

\begin{table*}
	\centering
	\caption{Physical Properties of the BAL Outflows and [\oiii] Emission Outflows}
	\label{tab:compare}
	\begin{tabular}{llllllll} 
		\hline
		\hline
		Object name					&J0046+0104			&J0825+0740		&J0831+0354		&J0941+1331		&J1111+1437		&J1135+1615		&J1512+1119$^{1}$ 	\\
		\hline
		\multicolumn{2}{l}{ (a) BAL Outflow Parameters: }\\
		$v$, km s$^{-1}$				& -1730				&+395 			&-10800   		&-3180			&-1860 			&-7250		   	&-1850 \\
		\civ\ Abs. width, km s$^{-1}$			& 3670$^{+40}_{-40}$		&2430$^{+40}_{-40}$ 	&4970$^{+40}_{-220}$   	&7980$^{+60}_{-60}$ 	&1770$^{+40}_{-40}$ 	&9500$^{+100}_{-40}$   &2190$^{+40}_{-40}$ \\
		log(\ne), cm$^{-3}$				& 3.8$^{+0.1}_{-0.1}$		&3.2$^{+0.2}_{-0.2}$ 	&4.4$^{+0.3}_{-0.2}$   	&4.0$^{+0.1}_{-0.1}$ 	&3.6$^{+0.1}_{-0.1}$ 	&$>$5.4$_{-0.3}$	&5.4$^{+2.7}_{-0.6}$ \\
		[0.5mm]
		log(\Uh), cm$^{-2}$				& -1.4$^{+0.2}_{-0.2}$		&-2.1$^{+0.4}_{-0.3}$ 	&-0.3$^{+0.5}_{-0.5}$   &-0.4$^{+0.2}_{-0.2}$ 	& -1.2$^{+0.2}_{-0.2}$	&-0.1$^{+0.2}_{-0.2}$ 	&-0.9$^{+0.1}_{-0.1}$ \\
		[0.5mm]
		log(\Nh), cm$^{-2}$				& 21.2$^{+0.2}_{-0.2}$		&20.4$^{+0.6}_{-0.4}$ 	& 22.5$^{+0.5}_{-0.4}$  &22.3$^{+0.2}_{-0.2}$	&21.5$^{+0.2}_{-0.3}$	&22.6$^{+0.2}_{-0.2}$  	&21.9$^{+0.1}_{-0.1}$ \\	
		[0.5mm]
		R$_{\text{distance}}$[BAL], pc			& 1200$^{+250}_{-450}$		&3900$^{+1100}_{-2800}$	&110$^{+30}_{-25}$   	&200$^{+40}_{-60}$ 	&880$^{+210}_{-260}$	&$<$40$^{+10}_{-40}$   	&10 -- 300\\	
		$\dot{M}$[BAL], $M_{\odot}$ yr$^{-1}$		& 37$^{+4}_{-3}$		&4.6$^{+4.0}_{-1.7}$   	&410$^{+530}_{-220}$ 	&120$^{+14}_{-13}$ 	&55$^{+10}_{-11}$	&$<$150$^{+10}_{-30}$   &1 -- 55 \\
		log(\Ek[BAL] in  erg s$^{-1}$)			& 43.5$^{+0.1}_{-0.1}$		&41.3$^{+0.27}_{-0.20}$ &46.2$^{+0.4}_{-0.3}$   &44.6$^{+0.04}_{-0.05}$	&43.8$^{+0.07}_{-0.1}$	&$<$45.4$^{+0.03}_{-0.1}$&43$^{+0.8}_{-0.9}$ \\
		\\
		\multicolumn{2}{l}{ (b) \hb\ + [\oiii] Gaussian fitting results: }\\
		\multicolumn{2}{l}{ For \hb\ emission line:}\\
		$v_{\text{c1}}$, km s$^{-1}$			& -150$^{+10}_{-10}$		&450$^{+60}_{-60}$ 	&420$^{+100}_{-100}$   	&0$^{+970}_{-970}$	&240$^{+20}_{-20}$ 	&5$^{+70}_{-70}$  	&-210$^{+40}_{-40}$ \\
		$\sigma_{\text{c1}}$, km s$^{-1}$		& 960$^{+8}_{-8}$		&1200$^{+50}_{-50}$ 	&350$^{+140}_{-149}$   	&1000$^{+3000}_{-3000}$ &970$^{+40}_{-40}$ 	&820$^{+60}_{-60}$  	&3600$^{+50}_{-50}$ \\
		$v_{\text{c2}}$, km s$^{-1}$			& -490$^{+40}_{-40}$		&-1300$^{+60}_{-60}$ 	&570$^{+109}_{-109}$   	&20$^{+310}_{-310}$ 	&1000$^{+120}_{-120}$ 	&4$^{+420}_{-420}$   	&-610$^{+10}_{-10}$ \\
		$\sigma_{\text{c2}}$, km s$^{-1}$		& 3500$^{+50}_{-50}$		&2600$^{+70}_{-70}$ 	&1760$^{+169}_{-169}$  	&2900$^{+1380}_{-1380}$ &3700$^{+160}_{-160}$ 	&4400$^{+2000}_{-2000}$ &250$^{+10}_{-10}$ \\
		\multicolumn{2}{l}{ For [\oiii] emission lines:}\\
		$v_{\text{c1}}$, km s$^{-1}$			& -430$^{+9}_{-9}$		&-7$^{+10}_{-10}$ 	&130$^{+90}_{-90}$   	&310$^{+20}_{-20}$ 	&130$^{+20}_{-20}$ 	&-850$^{+250}_{-250}$   &-360$^{+50}_{-50}$ \\
		$\sigma_{\text{c1}}$, km s$^{-1}$		& 200$^{+10}_{-10}$		&150$^{+20}_{-20}$ 	&310$^{+50}_{-50}$   	&150$^{+10}_{-10}$ 	&210$^{+30}_{-30}$ 	&1300 $^{+410}_{-410}$  &1700$^{+60}_{-60}$ \\
		$v_{\text{c2}}$, km s$^{-1}$			& -700$^{+30}_{-30}$		&-420$^{+80}_{-80}$ 	&-750$^{+70}_{-70}$   	&-80$^{+90}_{-90}$ 	&-330$^{+200}_{-200}$ 	&-1200$^{+250}_{-250}$  &-630$^{+1}_{-1}$ \\
		$\sigma_{\text{c2}}$, km s$^{-1}$		& 880$^{+30}_{-30}$		&330$^{+60}_{-60}$ 	&1100$^{+60}_{-60}$   	&570$^{+100}_{-100}$ 	&380$^{+110}_{-110}$ 	&350$^{+330}_{-330}$  	&190$^{+3}_{-3}$ \\
		$v_{\text{c3}}$, km s$^{-1}$			& -3000$^{+110}_{-110}$		&-2900$^{+60}_{-60}$ 	&-3500$^{+120}_{-120}$  &-2300$^{+620}_{-620}$ 	&-1200$^{+270}_{-270}$ 	&-2400$^{+1000}_{-1000}$&-670$^{+3}_{-3}$ \\
		$\sigma_{\text{c3}}$, km s$^{-1}$		& 990$^{+120}_{-120}$		&1500$^{+80}_{-80}$ 	&1100$^{+110}_{-110}$   &1300$^{+900}_{-900}$ 	&1100$^{+170}_{-170}$ 	&600$^{+640}_{-640}$   	&380$^{+10}_{-10}$ \\
		\\
		\multicolumn{2}{l}{ (c) [\oiii] measurements: }\\
		\multicolumn{2}{l}{ From Gaussian-fitted [\oiii] profiles: }\\		
		\vmed, km s$^{-1}$				& -770$^{+40}_{-40}$		&-2000$^{+140}_{-140}$ 	&-1400$^{+290}_{-290}$  &-310$^{+470}_{-470}$ 	&-210$^{+120}_{-160}$ 	&-1000$^{+360}_{-310}$  &-600$^{+20}_{-20}$ \\
		$w_{80}$, km s$^{-1}$				& 3200$^{+120}_{-120}$		&4500$^{+120}_{-80}$	&4500$^{+410}_{-410}$   &3600$^{+570}_{-2380}$ 	&2100$^{+250}_{-230}$ 	&3300$^{+1500}_{-1600}$ &1500$^{+20}_{-40}$ \\
		$w_{90}$, km s$^{-1}$				& 4200$^{+80}_{-101}$		&5200$^{+140}_{-140}$ 	&5300$^{+410}_{-410}$   &4400$^{+1060}_{-2900}$ &2800$^{+230}_{-230}$ 	&4200$^{+2100}_{-1900}$ &2900$^{+80}_{-80}$ \\
		$A$						& -0.40$^{+0.02}_{-0.01}$	&-0.11$^{+0.05}_{-0.05}$&-0.18$^{+0.1}_{-0.1}$	&-0.54$^{+0.3}_{-0.1}$	&-0.48$^{+0.1}_{-0.1}$	&0.04$^{+0.2}_{-0.2}$ &0.12$^{+0.04}_{-0.01}$ \\
		
		\multicolumn{2}{l}{ From non-parametric-fitted [\oiii] profiles: }\\
		\vmed, km s$^{-1}$				& -730$^{+10}_{-10}$		&-1490$^{+10}_{-10}$ 	&-1370$^{+10}_{-10}$  	&-210$^{+40}_{-40}$ 	&-720$^{+10}_{-10}$ 	&-1460$^{+50}_{-50}$  	&-540$^{+7}_{-7}$ \\
		$w_{80}$, km s$^{-1}$				& 3400$^{+10}_{-10}$		&4800$^{+20}_{-20}$	&4500$^{+10}_{-10}$   	&3400$^{+40}_{-40}$ 	&1900$^{+10}_{-10}$ 	&3600$^{+40}_{-40}$ 	&1400$^{+10}_{-10}$ \\
		$w_{90}$, km s$^{-1}$				& 4500$^{+10}_{-10}$		&5300$^{+20}_{-20}$ 	&5300$^{+10}_{-10}$   	&4000$^{+40}_{-40}$ 	&2500$^{+20}_{-20}$ 	&4800$^{+40}_{-40}$ 	&2700$^{+10}_{-10}$ \\
		$A$						& -0.43$^{+0.01}_{-0.01}$	&-0.11$^{+0.01}_{-0.01}$&-0.19$^{+0.01}_{-0.01}$&-0.56$^{+0.02}_{-0.02}$&-0.45$^{+0.01}_{-0.01}$&0.05$^{+0.10}_{-0.10}$ &0.10$^{+0.01}_{-0.01}$ \\
		\\
		
		\multicolumn{2}{l}{ (d) Luminosities and Energetics}\\
		log(L$_{\text{bol}}$ in erg s$^{-1}$)		& 47.1$^{+0.03}_{-0.04}$	&46.8$^{+0.07}_{-0.08}$	&47.00$^{+0.04}_{-0.01}$&46.8$^{+0.02}_{-0.06}$	&46.9$^{+0.02}_{-0.10}$	&47.2$^{+0.02}_{-0.03}$	&47.6$^{+0.01}_{-0.03}$ \\
		log(L[\oiii] in erg s$^{-1}$)			& 43.9$^{+0.02}_{-0.02}$	&43.8$^{+0.02}_{-0.02}$	&43.70$^{+0.01}_{-0.01}$&43.2$^{+0.2}_{-0.4}$	&43.7$^{+0.09}_{-0.11}$	&43.3$^{+0.1}_{-0.4}$	&44.3$^{+0.01}_{-0.01}$ \\
		log(L[\hb] in erg s$^{-1}$)			& 44.7$^{+0.01}_{-0.01}$	&44.3$^{+0.01}_{-0.01}$	&43.70$^{+0.04}_{-0.02}$&43.90$^{+0.2}_{-0.2}$	&44.4$^{+0.02}_{-0.02}$	&44.0$^{+0.1}_{-0.3}$	&44.7$^{+0.01}_{-0.01}$ \\
		log($\nu$L$_{\nu}$[1450\angstrom] in erg s$^{-1}$)&46.0$^{+0.03}_{-0.05}$	&45.6$^{+0.05}_{-0.04}$	&45.70$^{+0.04}_{-0.04}$&45.7$^{+0.02}_{-0.02}$	&45.7$^{+0.04}_{-0.08}$	&46.0$^{+0.01}_{-0.02}$	&46.0$^{+0.04}_{-0.07}$ \\
		R$_{\text{size}}$[\oiii], pc			& 520$^{+10}_{-30}$		&1600$^{+100}_{-100}$ 	&55$^{+2}_{-3}$   	&100$^{+2}_{-2}$ 	&400$^{+20}_{-20}$ 	&19$^{+0.4}_{-0.4}$   	&50$^{+2}_{-2}$ \\	
		\hline
	\end{tabular}
	\\ [5mm]
	\raggedright
	
	\textbf{Table 2.}\\
	(a) The detemined physical parameters for the observed BAL outflow are reported in \cite{Borguet12b, Chamberlain15b, Xu18}, and \cite{Xu19a}. See a summary in table 4 of in \cite{Xu19a}.\\
	(b) Fitting results for the \hb\ + [\oiii] region. All velocity centroids ($v_{\text{c}}$) and widths ($\sigma_{\text{c}}$) are relative to the quasar's rest-frame. The errors represent the 1$\sigma$ uncertainties of the fitting, which are computed from the covariance matrix in the fitting processes (see section \ref{sec:OIII} for details).\\
	(c) The [\oiii] measurements and their corresponding errors of the observed [\oiii] features from two different approaches, i.e., Gaussian fitting approach (section \ref{sec:OIII}) and non-parametric fitting approach (section \ref{sec:nonpara}).\\
	(d) The luminosities and energetics of the [\oiii] outflows where more details are in section \ref{sec:discuss}.\\
	$^{1}$ We show here the outflow system 2 of SDSS J1512+1119 \citep{Borguet12b}.\\
	\textbf{Notes of Abbreviations:}\\
	(a): \ne: electron number density, \Uh: ionization parameter, \Nh: total hydrogen column density, R$_{\text{distance}}$[BAL]: derived distance of the BAL outflow to the central AGN, $\dot{M}$: outflow's mass flow rate, \Ek: kinetic luminosity of the outflow.\\
	(b): $v_{\text{c1}}$ --  $v_{\text{c3}}$: velocity centroids of different Gaussian components, $\sigma_{\text{c1}}$ --  $\sigma_{\text{c3}}$: velocity widths of different Gaussian components (section \ref{sec:OIII}).\\ 
	(c): \vmed, $w_{80}$, $w_{90}$, and $A$ are defined in section \ref{sec:nonpara1}.\\
	(d): L$_{\text{bol}}$: bolometric luminosity of the quasar, L[\oiii]: observed [\oiii] luminosity, L[\hb]: observed \hb\ luminosity, $\nu$L$_{\nu}$[1450\angstrom]: observed continuum luminosity at 1450\angstrom\ rest-frame, R$_{\text{size}}$: size of the total illuminating area of the [\oiii] emission region (section \ref{sec:energy}).


\end{table*}

\subsubsection{[\oiii] Measurements}
\label{sec:nonpara1}

From the [\oiii] emission fits in section \ref{subsec:OIII}, we measure the line properties by calculating the accumulated line flux at each velocity \citep{Whittle85a}. The cumulative [\oiii] \ly 5007 flux as a function of velocity is:
\begin{equation}
    \Phi(v) = \int_{-\infty}^{v} F(v') dv'
    \label{func:phiv}
\end{equation}
where we use the fitted [\oiii] profile from section \ref{subsec:OIII} as F(v$'$). These profiles are shown as the solid black lines in the panel (c) of figure \ref{fig:J0046all}. Adopting other fitting functions or adding more Gaussian components has minimal effects on the derived [\oiii] profile and measurements \citep[see discussion in section 2.3 of][and section \ref{sec:nonpara} below]{Zakamska14}. 

From equation \ref{func:phiv}, the total [\oiii] line flux is given by $\Phi(\infty)$, and v$_{a}$ represents the velocity when $\Phi$(v$_{a}$) = $\frac{a}{100}$ $\times \Phi(\infty)$ (0\% $<$ a $<$ 100\%). As an example shown in the panel (c) of figure \ref{fig:J0046all}, we plot the locations of $v_{05}$, $v_{10}$, $v_{50}$, $v_{90}$, and $v_{95}$ as purple lines from left to right, respectively (with remaining objects in Appendix \ref{sec:APPA}). The median velocity (\vmed\ $\equiv$ $v_{50}$) for the seven objects in our sample is in a range between -2000 and -210 km s$^{-1}$ (see table \ref{tab:compare}). Therefore, the observed [\oiii] emission features are strongly blueshifted relative to the \hb\ emission line centroids. The velocity widths of [\oiii] \ly 5007 are given by $w_{90} \equiv v_{95} - v_{05}$, and is in a range between 2800 and 5200 km s$^{-1}$. Such [\oiii] emission feature widths are similar to that of the ERQs reported in \cite{Zakamska14} and \cite{Perrotta19} (see discussion in section \ref{sec:kine}). $w_{90}$ is the velocity width of the emission between $v_{05}$ and $v_{95}$, and is shown as the gray regions in the panel (c) of figure \ref{fig:J0046all}. To quantify the line symmetry of the emission features, an asymmetry parameter is defined as \citep[e.g.,][]{Whittle85a, Liu13b}:
\begin{equation}
    A = \frac{(v_{90} - v_{\text{med}}) - (v_{\text{med}} - v_{10})} {w_{80}}
\end{equation}
For all 7 objects in the sample, we get $A$ ranging between --0.54 and 0.12. The negative $A$ parameter indicates a blue-shifted emission profile. We summarize the derived parameters for all 7 objects in part (c) of table \ref{tab:compare}. To estimate the errors on $A$, we assume the best fitting Gaussian parameters from section \ref{sec:OIII} are uncorrelated. Then we vary each Gaussian parameter in a range of $\pm$ 1$\sigma$ and calculate the resulting changes in $A$. We adopt the resulting maximum and minimum values of $A$ as its upper and lower limit, respectively. Similarly, we estimate the errors for the other derived parameters, including $v_\text{med}$, $w_{80}$, and $w_{90}$. 

\subsection{Non-parametric Fittings to the [\oiii] Emission Features}
\label{sec:nonpara}
In section \ref{sec:OIII}, we fit the [\oiii] emission features with 2 -- 3 Gaussian profiles, but in principle, other fitting functions (e.g., Lorentzian profiles) could be used. Moreover, there is no particular physical interpretation to the parameters of each Gaussian component. \cite{Villforth18} introduced a non-parametric fitting approach, which adopts the observed emission profile as a free function to fit the [\oiii] region. This method does not assume the [\oiii] emission's shape and instead takes as input only the observed spectral shape of [\oiii], the emission line strength ratio (1:2.99 between [\oiii] \ly 4959 and \ly 5007) and the wavelength difference between [\oiii] \ly 4959 and [\oiii] \ly 5007. The fitting is performed by decomposing the [\oiii] emission profiles into two tightly related functions for [\oiii] \ly 4959 and \ly 5007. One is for the [\oiii] \ly 5007 emission, and the other is for the [\oiii] \ly 4959 emission shifted by $\sim$ 48\angstrom\ and multiplied by 1/2.99 from the [\oiii] \ly 5007 function \citep[see equations (1) and (2) in][]{Villforth18}.





We adopt the same approach to fit the [\oiii] emission profiles for each object in our sample. We show the best fitting results for quasar SDSS J0046+0104 in the panel (d) of figure \ref{fig:J0046all}. The [\oiii] \ly 5007 profile obtained from the non-parametric approach (blue lines) is consistent with the one from the Gaussian fitting approach (solid blue lines in panel (b)). In order to compare the derived [\oiii] measurements between these two approaches, we again measure the [\oiii] properties, including \vmed, $w_\text{80}$, $w_\text{90}$, and $A$. The derived values are shown in part (c) of table \ref{tab:compare}. These two different fitting approaches give consistent [\oiii] measurements within error bars for most of the objects. Hereafter, we adopt the derived [\oiii] measurements based on the Gaussian-fitted [\oiii] profiles since most previous studies in the literature adopted the same methodology.\\

\section{Relationships between the BAL outflows and [\oiii] emission outflows}
\label{sec:connect}

In our sample, all seven objects present BAL troughs from various doublet transitions, e.g., \civ\ \ly\ly 1548.19, 1550.77, \nv\ \ly\ly 1238.82, 1242.80, \ovi\ \ly\ly 1031.93, 1037.62, \siiv\ \ly\ly 1393.75, 1402.77, and \siv\ \ly 1062.66 \citep{Borguet12b, Chamberlain15b, Xu18, Xu19a}. We also observe [\oiii] emission in all objects (see section \ref{sec:analysis}). To explore the relationships between these two types of outflow features, we first introduce how we measure the physical properties of BAL outflows in section \ref{sec:PPBAL}. Then we present the relationship between \ne\ (and $R$) derived from the BAL outflows and the luminosity of the [\oiii] emission features in section \ref{sec:neVSLum}. Following that, we compare the velocity widths between the BAL and emission outflows in section \ref{sec:compV}, and we calculate and compare the average radial velocity of the two outflows in section \ref{sec:compAverV}. The unique patterns and notes on individual object are discussed in section \ref{sec:note}. Finally, we summarize these relationships in section \ref{sec:sumCon}.\\

\subsection{Measuring the Physical Properties of BAL Outflows}
\label{sec:PPBAL}

Here we give a brief description for the methods we used to extract the physical properties of the BAL outflows in our sample. Full discussion can be found in \cite{Borguet12b, Chamberlain15b, Xu18,Xu19a}.

1. We begin with measuring the ionic column densities from the observed absorption troughs \citep[section 4.2 of ][]{Xu19a}, where we used the apparent optical depth (AOD), partial-covering, and power-law absorption models \citep[e.g.,][]{Arav05}. 

2. Ionization equilibrium in an AGN outflow is dominated by photoionization \citep[e.g.,][]{Arav01}, which depends on the ionization parameter (\Uh):
\begin{equation}
\label{Eq:Uh}
\Uh=\frac{\Qh}{4\pi R^2 \nh c}
\end{equation}
where $\Qh$ is the source emission rate of hydrogen ionizing photons (obtained by integrating the specific flux, $F_{\nu}$, for energies above 1 Ryd), $R$ is the distance of the outflow from the central source, $\nh$\ is the hydrogen number density (for a highly ionized plasma, \nh\ $\simeq$ 0.8 \ne), and c is the speed of light. In order to derive $R$, we need determinations of \Uh\ and \ne.

3. To determine \Uh, we perform photoionization analysis by running the spectral synthesis code Cloudy [version c17.00, \cite{Ferland17}] to generate a grid of photoionization simulations. We compare the model predictions and measured column densities in the phase space of \Uh\ and total hydrogen column density, \Nh. The best-fitting photoionization solution (i.e., \Uh\ and \Nh) for the outflow is found through $\chi^{2}$-minimization of the difference between the model-predicted and the measured column densities \citep[see figure 4 in ][]{Xu19a}. 

4. Since the \siv*\ energy level is populated by collisional excitations from free electrons, the ratio between the column densities of \siv*\ \ly 1072.97 and \siv\ \ly 1062.66 can be used as a diagnostic for \ne. We measure the column densities for the absorption troughs of these two transitions and compare with the theoretical predictions from CHIANTI \citep[version 7.1.3, ][]{Landi13}. The determined \ne\ for all 7 objects in our sample are given in table \ref{tab:compare} here.

5. With the knowledge of \Uh\ and \ne, we can solve for $R$ using equation (\ref{Eq:Uh}). The derived $R$ values for the BAL outflows are summarized in table 4 of \cite{Xu19a}.

6. Assuming the outflow is in the form of a thin shell \citep[][]{Borguet12a}, the mass flow rate ($\dot{M}$) and kinetic luminosity (\Ek) of the observed BAL outflows in our sample are calculated by: \\
\begin{equation}\label{eq:dotM}
\begin{split}
\dot{M}\simeq 4\pi \Omega R\Nh \mu m_p v 
\end{split}
\end{equation}

\begin{equation}
\begin{split}
\dot{E}_{k}\simeq \frac{1}{2} \dot{M} v^2 
\end{split}
\label{eq:Edot}
\end{equation}
where $\mu$ = 1.4 is the mean atomic mass per hydrogen atom, $m_p$ is the proton mass, $\mathit{v}$ is the velocity of the BAL outflow, and $\Omega$ = 0.08 is the global covering factor for BAL outflows which exhibit \siv\ and \siv* absorption troughs \citep[see discussion in section 5.2 in][]{Borguet13}. The derived physical properties for the BAL outflows in our sample are shown in part (a) of table \ref{tab:compare} \citep[][]{Borguet12b, Chamberlain15b, Xu18, Xu19a}.



\subsection{Density/Distance of BAL outflows and [\oiii] Luminosity}
\label{sec:neVSLum}

\begin{figure}
	\includegraphics[width=1.0\columnwidth,angle=0,trim={0.5cm 0.5cm 5.5cm 3cm}]{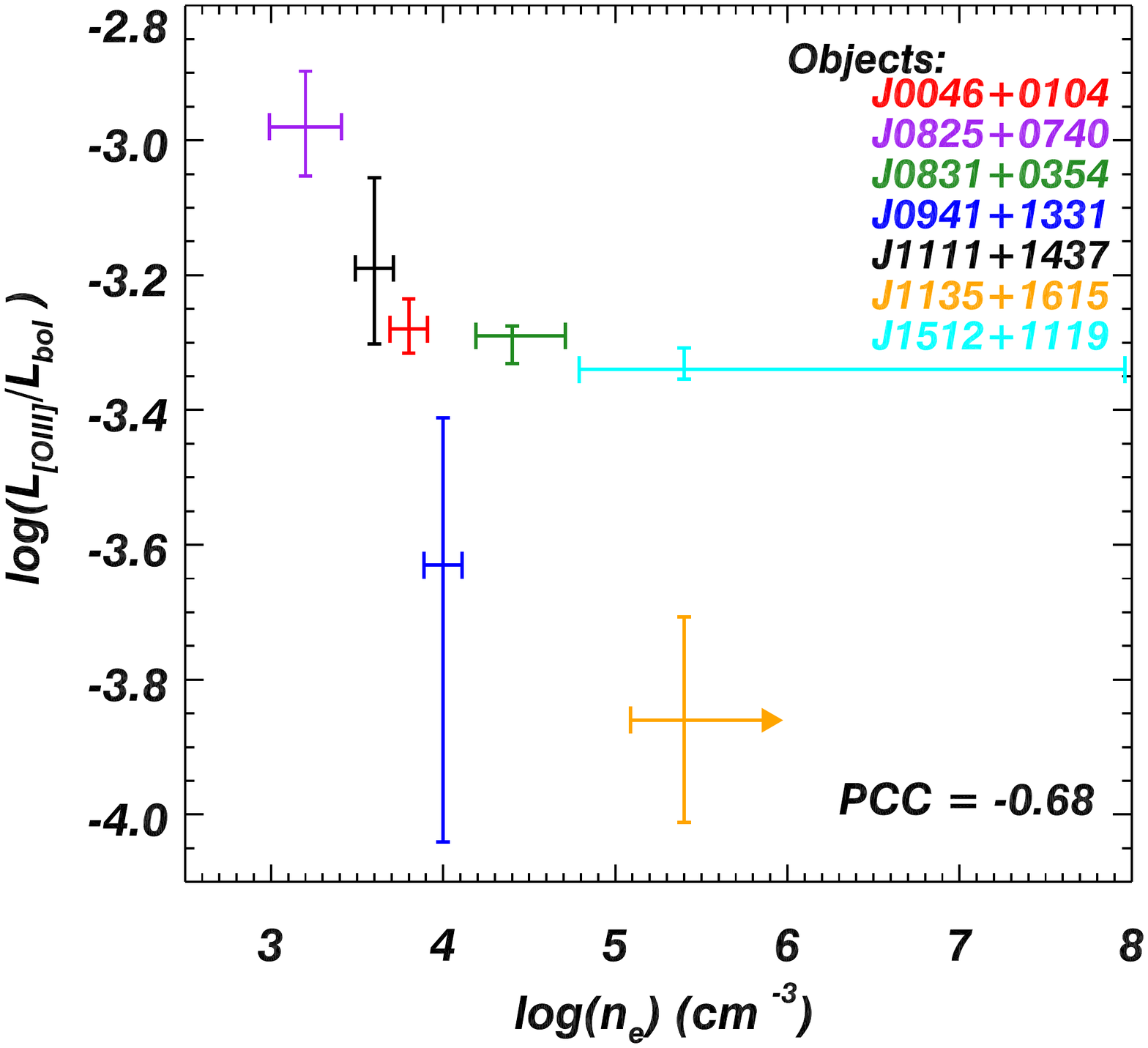}
	\includegraphics[width=1.0\columnwidth,angle=0,trim={0.5cm 1cm 5.5cm 2cm}]{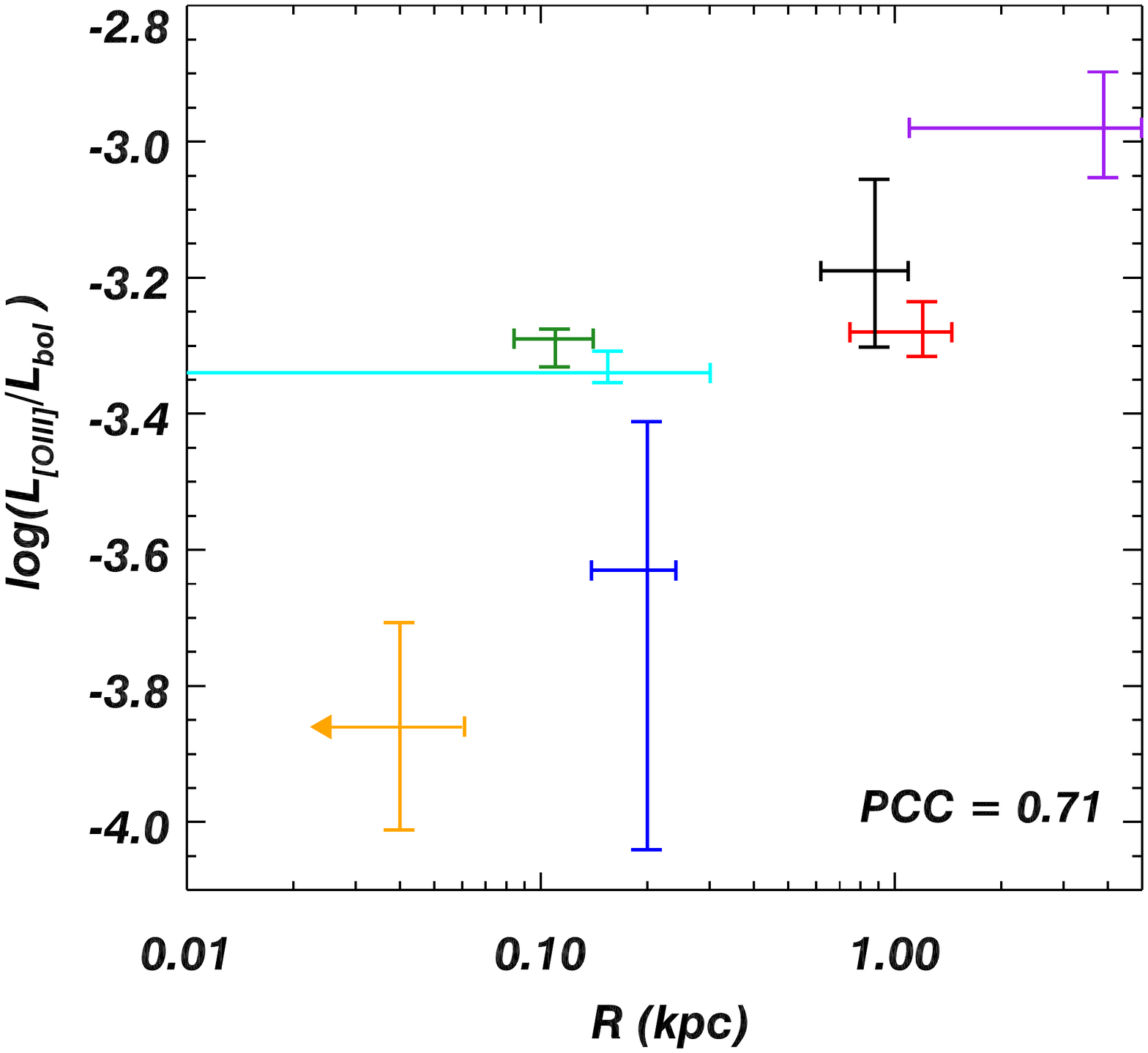}
    \caption{\textbf{Top}: [\oiii] luminosity normalized by the quasar's bolometric luminosity (L$_\text{[OIII]}$/L$_\text{bol}$) versus the electron number density (\ne) derived for the BAL outflows. The values and corresponding errors are shown and color coded for different objects. The Pearson correlation coefficient (PCC) between the x- and y-values is shown at the bottom-right corner, where 1 is total positive linear correlation, 0 is no linear correlation, and --1 is total negative linear correlation. \textbf{Bottom}: L$_\text{[OIII]}$/L$_\text{bol}$ versus the reported BAL outflow distance ($R$) (the values and the corresponding errors are given in table \ref{tab:compare} and references therein). The \civ\ outflow in quasar SDSS J1135+1615 has a lower limit on \ne, and therefore an upper limit on $R$. We show these limits as orange arrows. }
    \label{fig:compne}
\end{figure}

\begin{figure}
	\includegraphics[width=1.0\columnwidth,angle=0,trim={0.5cm 1.5cm 5.5cm 3cm}]{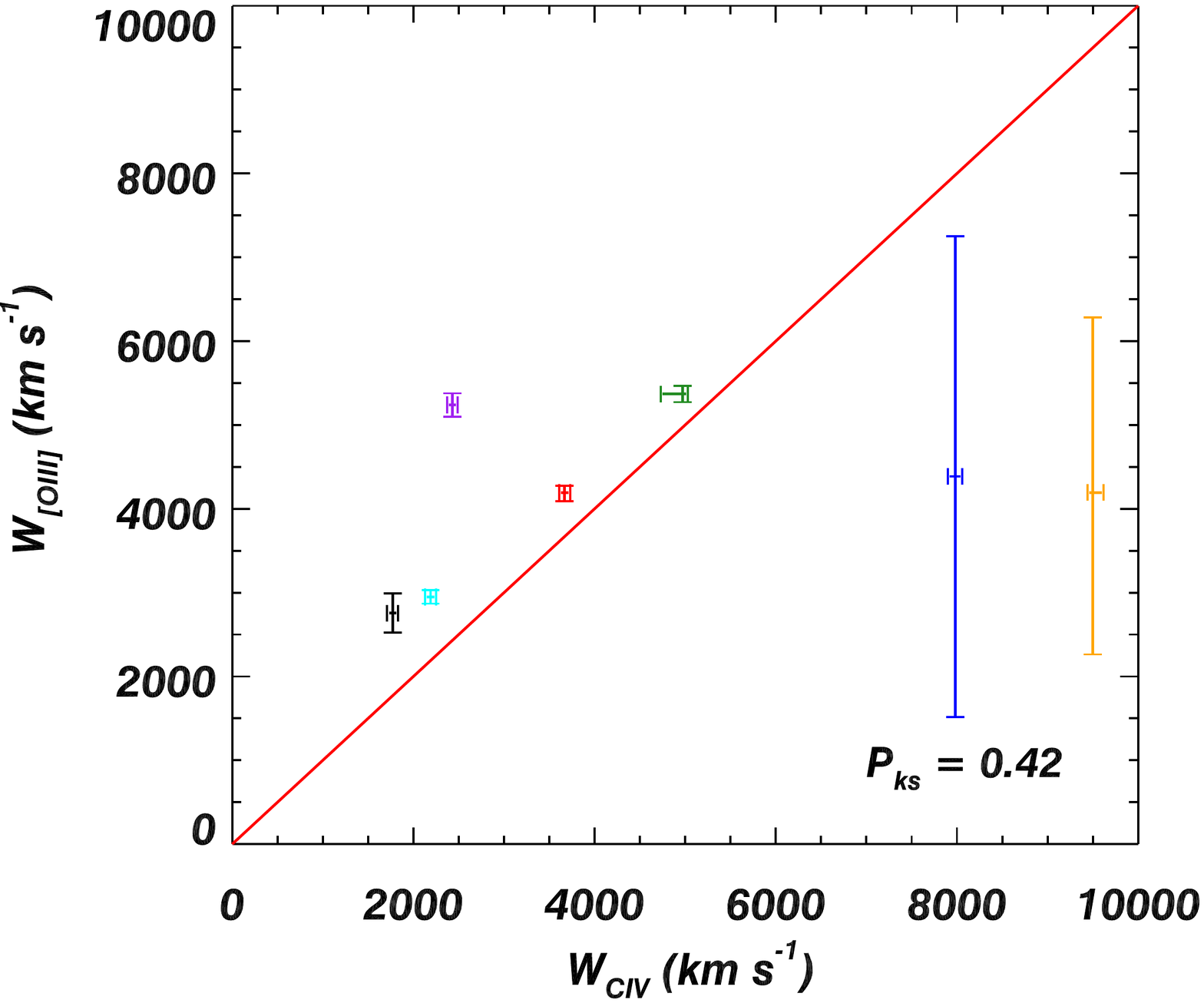}
	\includegraphics[width=1.0\columnwidth,angle=0,trim={0.5cm 2cm 5.5cm 1cm}]{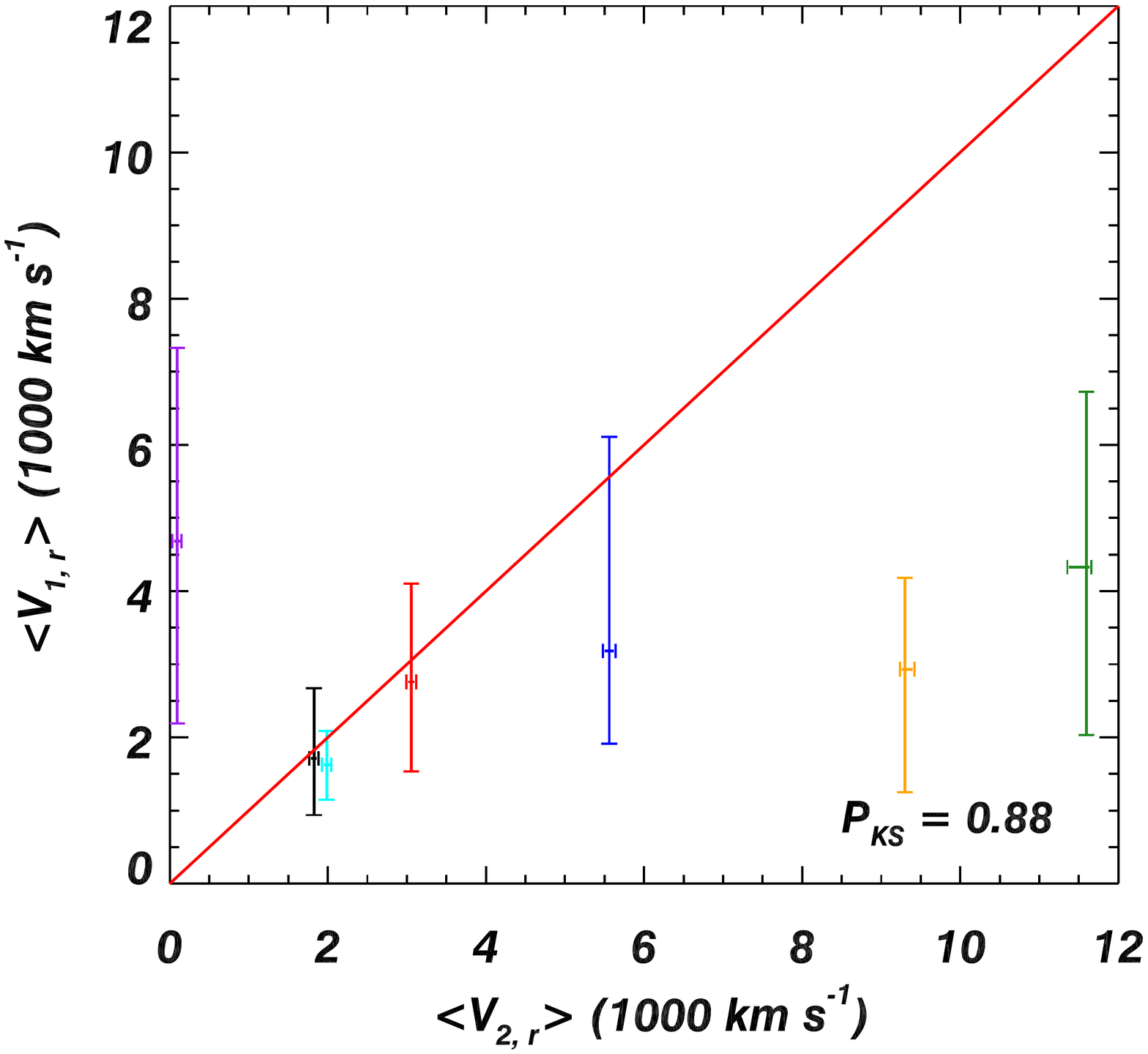}
    \caption{\textbf{Top:} Comparison between $w_{90}$ from [\oiii] emission features and the \civ\ absorption trough widths in our sample. The values and corresponding errors are shown and color coded for different objects (same as figure \ref{fig:compne}). See discussion in section \ref{sec:compV}. \textbf{Bottom:} Comparison of the mean radial velocity between the [\oiii] emission (\vro) and \civ\ absorption (\vrc) outflows in our sample. See section \ref{sec:compAverV} for details. In both panels, the red lines have slopes as unity and the color patterns are the same as figure \ref{fig:compne}. For each panel, the p value from the Kolmogorov–Smirnov (P$_{ks}$) test between the x- and y-axis quantities is shown at the bottom-right corner. A large P$_{ks}$ value means that the x- and y-axis quantities are consistent with being drawn from the same distribution.}
    \label{fig:compV}
\end{figure}

As stated in section \ref{sec:intro}, low gas densities (n $\lesssim$ 7 $\times$ 10$^{5}$ cm$^{-3}$) are needed for generating [\oiii] emission features \citep{Baskin05}. From IFU observations, the surface brightness of [\oiii] emission peaks in the central regions of quasars and declines, but can be detected out to many kilo-parsecs from the center \citep[e.g.,][]{Liu13a, Liu13b, Harrison14}. The BAL density and distance information allows us to test the physical similarity between the emission and absorption outflows detected in the same quasar.



We observed strong blue asymmetries in most of the measured [\oiii] \ly 5007 profiles, which suggest that the [\oiii] emission profiles are affected by extinction. If we assume that the [\oiii] emission features originate in a thin spherical shell \citep[][and see an example in figure \ref{fig:OIIIgeo}]{Borguet12a}, the extinction would mostly affect the red-shifted portion of the emission \citep[e.g.,][]{Zakamska16, Perrotta19}. Therefore, we estimate the luminosity of the [\oiii] emission (L$_\text{[OIII]}$) by the blue-shifted portion of the emission profile and normalize it by the bolometric luminosity of the quasar \citep[L$_\text{bol}$, see table 2 in][]{Xu19a}. These values and the corresponding errors are shown in part (d) of table \ref{tab:compare}. In the top panel of figure \ref{fig:compne}, we show the derived L$_\text{[OIII]}$/L$_\text{bol}$ versus \ne\ for the BAL outflow in the same object. Each object's value and the corresponding errors are shown as the colored labels. Similarly, in the bottom panel of figure \ref{fig:compne}, we show L$_\text{[OIII]}$/L$_\text{bol}$ versus $R$ for the BAL outflow in the same object. 

Figure \ref{fig:compne} shows two trends: 1) L$_\text{[OIII]}$/L$_\text{bol}$ tends to decrease with increasing \ne\ (top panel) and 2) L$_\text{[OIII]}$/L$_\text{bol}$ also tends to increase with increasing $R$ (bottom panel), where $R$ is derived from \ne\ using equation \ref{Eq:Uh}. The BAL outflow in quasar SDSS J1135+1615 has a lower limit on \ne\ ($>$ 10$^{5.4}$ cm$^{-3}$, and is consistent with arbitrarily higher \ne) and therefore an upper limit on $R$ ( $<$ 40 pc). The observed L$_\text{[OIII]}$/L$_\text{bol}$ for quasar SDSS J1135+1615 is the smallest compared to the ones in the other objects. In contrast, the BAL outflow in quasar SDSS J0825+0740 has the smallest \ne\ (10$^{3.2}$ cm$^{-3}$) and the largest $R$ (3.9 kpc) with the highest L$_\text{[OIII]}$/L$_\text{bol}$. The observed trends suggest the BAL-derived densities apply to the [\oiii]-emitting region. Therefore, the anti-correlation between \ne\ and [\oiii] luminosity would be indicative of collisional suppression of [\oiii] at highest \ne\ and smallest $R$.




\subsection{Velocity Widths Comparison}
\label{sec:compV}

In the top panel of figure \ref{fig:compV}, we compare the widths between the outflows obtained from each quasar, i.e., [\oiii] emission outflow's width (\WOIII) indicated by $w_{90}$ and BAL outflow's width (\WCIV) measured for continuous \civ\ absorption below the normalized flux I = 0.9. The seven objects in our sample are shown in colored labels with the corresponding errors (the same color patterns as figure \ref{fig:compne}). The solid red line indicates the positions where \WOIII\ = \WCIV. Based on the assumed geometry in figure \ref{fig:OIIIgeo}, the observed [\oiii] emission could originate from different places. Therefore, there are two main contributions to \WOIII: 1) the angle between the radial velocity of the outflowing gas and the line-of-sight (LOS) direction, 2) the velocity distribution within the outflowing gas which emits [\oiii]. For \WCIV, the observed absorption troughs are formed within the LOS, and therefore, the velocity distribution of the outflowing gas is the main contribution. 

From our sample, the relationship between \WOIII\ and \WCIV\ is unclear, but these two widths in the same object are similar (within a factor of 3). Four objects show higher \WOIII\ than \WCIV\ (top panel of figure \ref{fig:compV}). The two objects which have lower \WOIII\ than \WCIV\ are quasar SDSS J0941+1331 (blue) and SDSS J1135+1615 (orange), where their \WOIII\ have large error bars. The p value (P$_{ks}$) from the Kolmogorov–Smirnov (KS) test between \WOIII\ and \WCIV\ is 0.42. This large P$_{ks}$ value indicates that the the distributions of \WOIII\ and \WCIV\ are consistent with being drawn from the same population.

\subsection{Radial Velocities Comparison}
\label{sec:compAverV}

We estimate the radial velocity of the outflowing gas ($v_\text{r}$) with the observed velocity profiles following the same methodology described in section 4.2 of \cite{Zakamska16}. As discussed in section \ref{sec:neVSLum}, the strong blue asymmetries in most of the observed [\oiii] emission profiles suggest that they are affected by extinction. We assume that 1) the extinction mostly affects the redshifted portion of the [\oiii] emission profile, 2) the [\oiii] outflow has a radial velocity distribution, $f$(\vro)d\vro, where $f(\vro)$ is proportional to the luminosity emitted by the [\oiii] gas with velocities between \vro\ and \vro\ + d\vro, and 3) the [\oiii] outflow has spherical symmetry in the hemisphere towards the observer \citep[e.g.,][]{Perrotta19}. Under these assumptions, regardless of the form of $f(\vro)$, we have \citep{Zakamska16}:

\begin{equation}
    \left \langle v_\text{$\scriptstyle{1}$,r}\right \rangle = 2 \left \langle |v_\text{$\scriptstyle{1}$,z}|\right \rangle
    \label{func:vr}
\end{equation}
where $\left \langle \vro \right \rangle$ is the mean radial velocity of the [\oiii] emission outflow, and $\left \langle |\vzo|\right \rangle$ is the flux-weighted average of the observed LOS velocity of [\oiii], \vzo, on the blushifted portion of the emission profile, i.e.,

\begin{equation}
    \left \langle |v_\text{$\scriptstyle{1}$,z}|\right \rangle =  \left | \frac{\int_{-\infty}^{0}v_\text{$\scriptstyle{1}$,z} \times F(v_\text{$\scriptstyle{1}$,z}) dv_\text{$\scriptstyle{1}$,z}}{\int_{-\infty}^{0} F(v_\text{$\scriptstyle{1}$,z}) dv_\text{$\scriptstyle{1}$,z}} \right | 
\label{func:mean_vz}
\end{equation}
where $F(\vzo)$ is the observed flux at velocity \vzo. Once we calculate $\left \langle |\vzo|\right \rangle$, we double the value and treat it as $\left \langle \vro \right \rangle$. Since obscuration more strongly affects streamlines which are further away from the LOS, the flux from streamlines which are close to parallel to the sky may be underestimated. Thus, $\left \langle \vro \right \rangle$ estimated here is likely biased toward larger values \citep{Zakamska16}.

For the observed BAL outflows in our sample, since we measure them directly along the LOS, we have the mean radial velocity of the BAL outflow, i.e., $\left \langle \vrc \right \rangle$, equals their mean LOS velocity, i.e., $\left \langle |\vzc|\right \rangle$. Moreover, $\left \langle |\vzc|\right \rangle$ can be calculated similarly as equation (\ref{func:mean_vz}) with:

\begin{equation}
    \left \langle |v_\text{$\scriptstyle{2}$,z}|\right \rangle =  \left | \frac{\int_{}^{}v_\text{$\scriptstyle{2}$,z} \times \tau(v_\text{$\scriptstyle{2}$,z}) dv_\text{$\scriptstyle{2}$,z}}{\int_{}^{} \tau(v_\text{$\scriptstyle{2}$,z}) dv_\text{$\scriptstyle{2}$,z}} \right | 
\label{func:mean_vz2}
\end{equation}
where $\tau(\vzc)$ is the optical depth of the absorption trough at velocity \vzc\ and the integration range is the whole observed \civ\ absorption trough. We show in the bottom panel of figure \ref{fig:compV} the comparison between the calculated $\left \langle v_\text{r}\right \rangle$ of the emission and absorption outflows in our sample. Six out of the seven objects have larger $\left \langle v_\text{r}\right \rangle$ in the observed \civ\ outflow than the [\oiii] outflow, i.e., $\left \langle \vrc \right \rangle$ $\geq$ $\left \langle \vro \right \rangle$. Taking into account the fact that $\left \langle \vro \right \rangle$ is likely biased toward higher values, we conclude that most of the observed BAL outflows have larger mean radial velocities than the [\oiii] emission outflows. 

This observation is consistent with our assumed geometry in figure \ref{fig:OIIIgeo}. The BAL outflows will show the highest radial velocity (i.e.,~\vrc). The [\oiii] emission can arise from several places on the shell with the observed radial velocity \vrc\ $\geq$ \vro\ $\geq$ 0 due to orientation effects. The P$_{ks}$ value between $\left \langle \vzc\right \rangle$ and $\left \langle \vzo\right \rangle$ is 0.88. Similar to section \ref{sec:compV}, this large P$_{ks}$ value supports the null hypothesis of KS test, i.e., the distributions of $\left \langle \vzo\right \rangle$ and $\left \langle \vzc\right \rangle$ are consistent with being drawn from the same population.

\subsection{Notes on Individual Objects}
\label{sec:note}

\begin{itemize}
\item{\indent Quasar SDSS J0825+0740 has a ``red-shifted'' \siiv\ absorption troughs (see figure \ref{fig:OIII_all}). The redshift of this quasar (z = 2.204) is derived from the \mgii\ \ly 2800 emission line.  From \cite{Xu19a}, the \siiv\ outflow centered at +395 km s$^{-1}$ shows absorption troughs from multiple doublet transitions, e.g., \civ\ \ly\ly 1548.19, 1550.77, \nv\ \ly\ly 1238.82, 1242.80, \ovi\ \ly\ly 1031.93, 1037.62, and \siiv\ \ly\ly 1393.75, 1402.77, and excited transitions from \siv*\ \ly 1072.97. Therefore, it is suggested to be a ``red-shifted" quasar outflow instead of an intervening system \citep[e.g.,][]{Hall13, Reis18, Zhou19}. }


\item{\indent Quasar SDSS J1135+1615's BAL outflow has a lower limit on \ne, i.e., log(\ne) > 5.4 cm$^{-3}$, and therefore, an upper limit of R $<$ 40 pc \citep{Xu19a}. For the \hb\ + [\oiii] region, contaminations from atmospheric water absorptions and instrumental artifacts exist. However, we still observed a moderate amount of the [\oiii] emission (4950 -- 5000\angstrom\ rest-frame, see figure \ref{fig:A2}3). The derived [\oiii] parameters from this object have large error bars (see table \ref{tab:compare}).}


\item{\indent Quasar SDSS J1512+1119 shows a high-density BAL outflow, i.e., log(\ne) = 5.4$^{+2.7}_{-0.6}$ cm$^{-3}$, with $R$ = 10 -- 300 pc \citep{Borguet12b}. In this case, if the [\oiii] emission features have similar physical parameters (\ne \ and $R$) as the observed BAL outflows, we do not expect to observe strong [\oiii] emission. However, we observe strong, narrow, and symmetric [\oiii] emission features (see figure figure \ref{fig:A2}2). This is not surprising since \cite{Zakamska14} found that broader Gaussian components of [\oiii] emission features are the ones that tend to be more asymmetric relative to the narrow ones (see their section 3.2). The absence of the broader component of [\oiii] emission in quasar SDSS J1512+1119 is consistent with the fact that the high \ne\ derived from the BAL outflows in this object produces minimal asymmetric [\oiii] emission.}

\end{itemize}


\begin{figure*}
	\includegraphics[width=1.0\columnwidth,angle=0,trim={0.5cm 1.5cm 7.cm 3cm}]{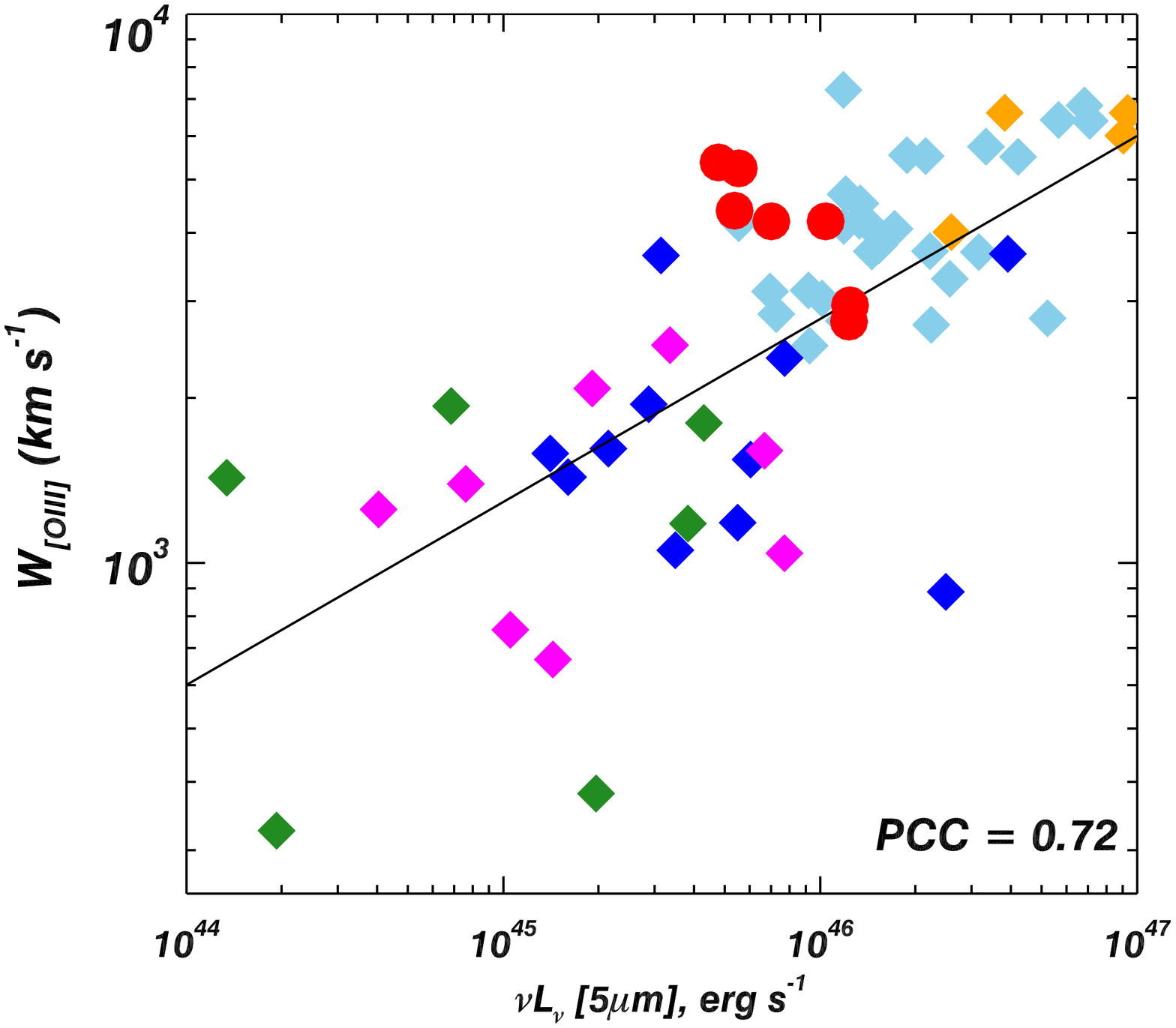}
	\includegraphics[width=1.0\columnwidth,angle=0,trim={0.5cm 1.5cm 7.cm 3cm}]{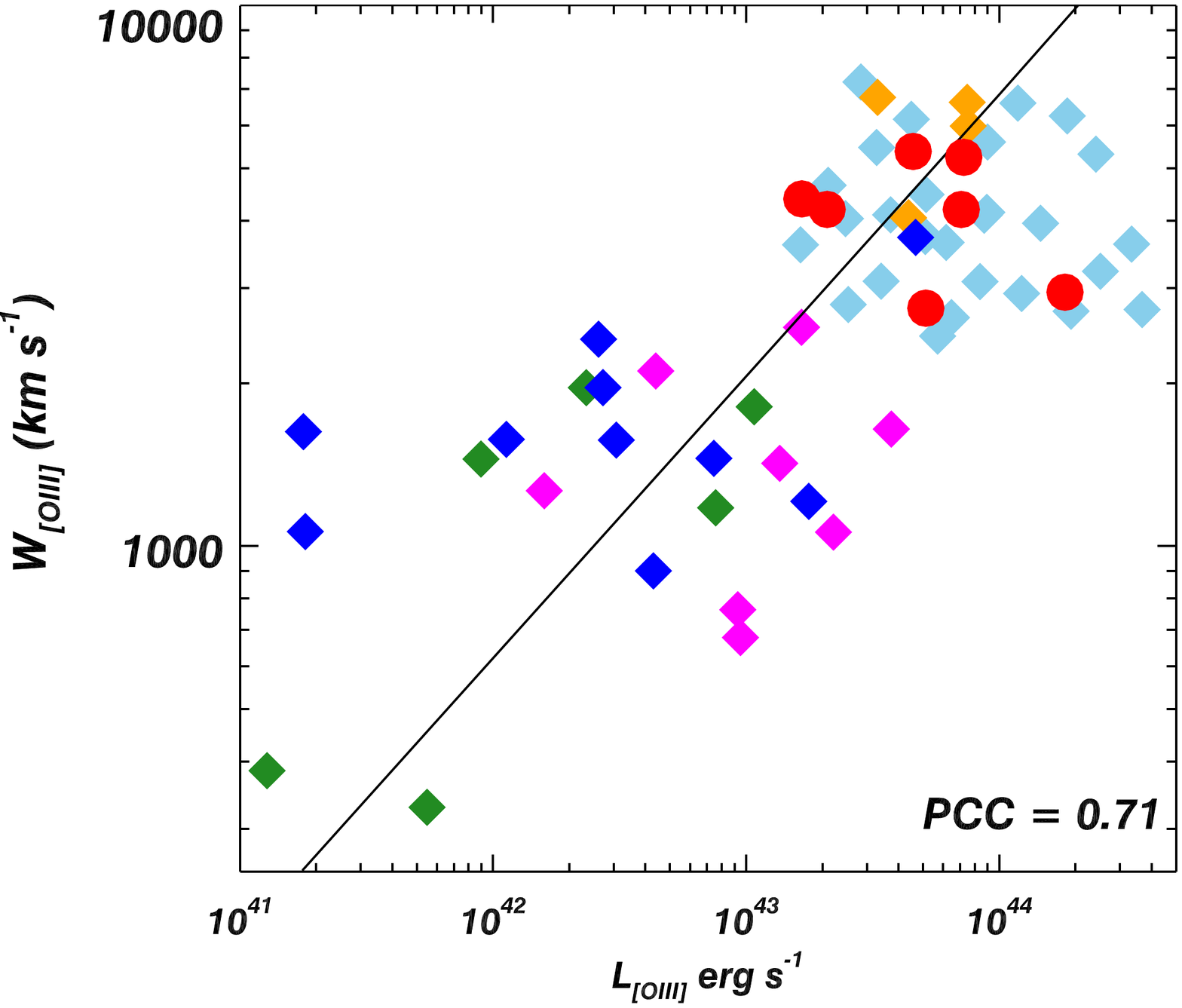}
    \caption{ [\oiii] kinematics as a function of the quasar's mid-infrared luminosities \textbf{(Left)} and [\oiii] luminosities \textbf{(Right)}. The width of [\oiii] emission feature (W$_\text{[OIII]}$) is measured by $w_{90}$, see section \ref{sec:nonpara1}). The objects in our sample are shown as the red circles. We compare with several quasar samples, including 28 extreme red quasars (ERQs) or ``ERQ-like" quasars in light-blue diamonds \citep{Perrotta19}, four objects from a optical+near-infrared selected ERQ sample shown as the orange diamonds \citep{Ross15, Zakamska16}; X-ray selected obscured quasars in green diamonds \citep{Brusa15a}; infrared-selected red quasars shown as blue diamonds \citep{Urrutia12}, and submillimeter-selected active galaxies shown as magenta diamonds \citep{Harrison12}. In both panels, the solid lines are the best-fitting relations between the [\oiii] kinematics and luminosities reported in section 4.1 of \protect\cite{Zakamska16}. Simliar to figure \ref{fig:compne}, the Pearson correlation coefficient (PCC) between the x- and y-values is shown at the bottom-right corner of each plot. See discussion in section \ref{sec:kine}.}

    \label{fig:comp1}
\end{figure*}

\subsection{Summary of the Relationships Between the Absorption and Emission Outflows}
\label{sec:sumCon}
To summarize, we find the following relationships between [\oiii] emission outflows and BAL outflows: 

1. In our sample, the observed normalized luminosity of [\oiii] (L$_\text{[OIII]}$/L$_\text{bol}$) is inversely proportional to the \ne\ derived from the BAL outflow in the same object. Similarly, L$_\text{[OIII]}$/L$_\text{bol}$ is proportional to $R$ from the BAL outflow in the same object (section \ref{sec:neVSLum}).

2. The measured velocity widths from the [\oiii] emission feature and the \civ\ absorption feature in the same object are similar (within a factor of 3, section \ref{sec:compV}).  

3. The mean radial velocity derived from most BAL outflows is larger than the one from the [\oiii] emission outflow in the same object (by a factor of 1 to 3, section \ref{sec:compAverV}). A KS test shows that the distributions of the [\oiii] and \civ\ radial velocities are consistent with being drawn from the same population.

4. The unique pattern in quasar SDSS J1512+1119 supports the claim that it is the broad component of the [\oiii] emission that connects more to the BAL outflows, instead of the narrow component.\\


\section{Discussions}

\label{sec:discuss}

\subsection{[\oiii] Properties and Comparisons to Other Samples}
\label{sec:kine}
As shown in section \ref{sec:intro}, strong and blue-shifted [\oiii] emission features in quasars, indicating high-velocity outflows, have been reported in the literature \citep[e.g.,][]{Zakamska14, Brusa15a, Perrotta19, Perna2019}. Here we compare the observed [\oiii] emission features in our sample to previous studies. 

In figure \ref{fig:comp1}, we plot the kinematics of [\oiii] (\WOIII, measured by $w_{90}$) as a function of the quasar's mid-infrared luminosity (left panel) and the [\oiii] luminosity (right panel). The objects from our sample are shown as red circles. The comparison samples are shown as colored diamonds. We estimate the mid-infrared luminosities at the 5 $\mu$m rest-frame following similar method described in \cite{Zakamska16}, i.e., a power-law that interpolates the closest observed near-infrared flux to the 5 $\mu$m rest-frame assuming $L_{\lambda}$ $\sim$ $\lambda^{-0.65}$ \citep[from the quasar spectral energy distribution in][]{Polletta07}.


In figure \ref{fig:comp1}, it is clear that most of our objects have a higher \WOIII\ than other samples (in green, blue, and magenta) and are comparable to those of the ERQs (in orange and light blue). A KS test of \WOIII\ from our sample and \cite{Perrotta19} yields that P$_{ks}$ = 0.38, which means the distributions of the two set of \WOIII\ are consistent with being selected from the same population. The objects in our sample are selected by their BAL outflow features with high luminosity, while the ERQs are selected based on their extreme infrared-to-optical or red-to-blue ratios \citep[e.g., section 2 of][]{Zakamska16}. The selection criteria of our BAL samples, even though quite different from the ERQs, clearly select objects with high [\oiii] widths. One possibility is that all quasars have BAL outflows, but those actually seen with BAL features are more likely to have [OIII] face-on. Another possibility is that only some quasars have BAL outflows and those are also more likely to have wide [OIII] outflows. 


In the left panel, combined with all previous studies, we observe a strong correlation between the [\oiii] kinematics and the quasar's mid-infrared luminosities \citep[see e.g.,][]{Zakamska14, Zakamska16, Perrotta19}. In the right panel, the observed [\oiii] luminosities for the objects in our sample are among the largest. This is not surprising since our BAL samples are selected with high bolometric luminosity (L$_\text{bol}$ $\gtrsim$ 10$^{47}$ erg s$^{-1}$, and see table \ref{tab:compare}).

\subsection{Possible Geometry For the BAL and Emission Outflows}
\label{sec:geometry}
\begin{figure}
	\includegraphics[width=0.92\columnwidth,trim={3cm 4cm 3cm 4cm}]{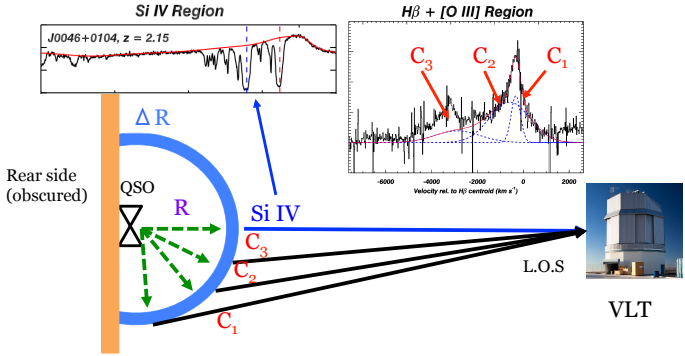}
    \caption{Sketch of a possible geometry where the same quasar outflow produces both a BAL and a blueshifted [OIII] emission signatures. BAL outflows are know to have small thickness ($\Delta R/R$ $\ll$ 1, e.g., Arav et al. 2013). For simplicity, we assume that the outflow is in form of a thin spherical shell, moving with velocity v$_{0}$. In this geometry, the BAL will show the highest radial velocity, while the [\oiii] emission can arise from several places on the shell with v$_{0}$ $>$ v$_{\text{[OIII]}}$ $>$ 0 as is the observed case. See section \ref{sec:geometry} for details.}
    \label{fig:OIIIgeo}
\end{figure}

A possible geometry for the two types of outflows is shown in figure \ref{fig:OIIIgeo}. We assume that the outflowing material is a thin spherical shell \citep[$\Delta$ R $\ll$ R, see][]{Arav13}, which covers a portion of the full solid angle around the quasar (blue circles). The line-of-sight (LOS) material creates the observed BAL outflows (blue solid lines). The material in all directions with suitable physical conditions \citep[i.e., relatively low \ne\ and warm (T $\sim$ 10$^4$ K) cloud, ][]{Zakamska16} creates the observed [\oiii] emission profiles with a spread of velocities due to orientation effects. The green arrows indicate the outflows radial velocity directions. We show the spectra of SDSS J0046+0104 as an example, where we adopted three Gaussians to fit its [\oiii] \ly 5007 profiles (c$_{1}$, c$_{2}$, and c$_{3}$, see table \ref{tab:compare}). These three velocity components of [\oiii] would originate from three different regions with different angles to the LOS.

\subsection{Physical Size of the [\oiii] Emission Region}
\label{sec:energy}
Following  equation (6) in \cite{Zakamska16}, the minimum size of [\oiii]-emitting region, R$_{s}$, can be estimated as:

\begin{equation}
    \begin{split}
        R_{s} = &0.5\ \text{kpc} \\
        &\times\ \left ( \frac{\nu L_{\nu} [1450\angstrom]}{10^{47}\ \text{erg s}^{-1}}\right ) ^{1/2}\ \left(\frac{\nh}{8 \times\ 10^5 \text{cm}^{-3}}\right )^{-1/2}\ \left (\frac{ U_\mathrm{\scriptstyle H} }{0.003}\right )^{-1/2}
    \end{split}
	\label{eq:rsize}
\end{equation}
where $\nu L_{\nu}$[1450\angstrom] is the quasar's luminosity at 1450\angstrom\ (rest-frame), \nh\ is the outflow's hydrogen number density, and \Uh\ is the outflow's ionization parameter [see equation (1)]. 


In our sample, the previously calculated densities and sizes of the BAL outflows are strongly correlated with the [\oiii] luminosity (section \ref{sec:connect}). The sense of the correlation is expected from the physics of the [\oiii] transition: those objects which have higher density/more compact BAL outflows appear to have a greater effect of collisional de-ionization and a lower [\oiii] luminosity. Moreover, we show in section \ref{sec:neVSLum} that the observed BAL outflows and the [\oiii] emission outflows have similar kinematics (i.e., $w_{90}$ and \vmed). These findings support the view that the observed emission and absorption outflows in the same object are associated with the same material (a possible geometry is shown in figure \ref{fig:OIIIgeo}). Therefore, we have a physical motivation to assume that the gas emitting the [\oiii] has similar \Uh\ and \ne\ to the BAL outflows in the same object. 

Adopting these values into equation (\ref{eq:rsize}), we calculate R$_{s}$ for the objects in our sample (see part (d) of table \ref{tab:compare}), which is in the range of 0.02 -- 1.60 kpc. The black hole masses for the objects in our sample are reported in table 2 of \cite{Xu19a} and are in the range of 10$^{8.8}$ -- 10$^9$ $M_{\odot}$. This leads to a broad line region size $\lesssim$ 0.004 kpc \citep[e.g.,][]{Kaspi05, Zakamska16}, which is much smaller than the derived R$_{s}$ values. The combination of BAL and [\oiii] analysis provide new estimates of the sizes of the [\oiii]-emitting region in quasars with high-velocity outflows (thousands of km s$^{-1}$). The derived R$_{s}$ values are consistent with IFU observations of [\oiii] emission regions \citep[e.g., up to 3 kpc from ][]{Liu13a, Liu13b}. These estimates can be tested with upcoming James Webb Space Telescope (JWST) observations \citep{Gardner06}.

\begin{table}

	\centering
	\caption{Physical Properties of the [\oiii] Emission Outflows}
	\label{tab:GlobalCovf}
	\scalebox{0.9}{
	\begin{tabular}{lllll} 
		\hline
		\hline
		Object name		&F[\oiii]$_\text{mod}^{(1)}$		&log(L[\oiii]$_\text{obs}^{(2)}$)	&S([\oiii]$^{(3)}$)		&$\Omega$([\oiii]$^{(4)}$		\\
					&(erg s$^{-1}$ cm$^{-2}$)		&log(erg s$^{-1}$)			&(cm$^{2}$)			&(\%)		\\
		\hline
		
		J0046+0104		&11 \   				&43.85					&6.1E42 		&3.6	   	\\
		J0825+0740		&0.5 \ \ \   				&43.86					&1.6E44 		&8.7		   	\\
		J0831+0354		&820 \ \   				&43.66					&5.6E40 		&3.8		\\		   	
		J0941+1331		&190 \ 					&43.22					&8.7E40 		&1.8	\\
		J1111+1437		&14 \   				&43.71					&3.7E42 		&4.0		\\
		J1135+1615		&7300 \  				&43.32					&2.8E39 		&1.5		\\
		J1512+1119		&2100   				&44.26					&8.6E40 		&3.0		\\
		\hline
	\end{tabular}
	}
	\\ [5mm]
	\raggedright
	Note. --\\
	(1): Model predicted flux of the [\oiii] emission from the best fitting photoionization solutions and \ne\ of the BAL outflows (section \ref{sec:covf}).\\
	(2): Observed luminosity of [\oiii] measured from the blue-shifted portion of the modeled [\oiii] \ly 5007 emission profiles (section \ref{sec:neVSLum})\\
	(3): Total illuminating area of the [\oiii] emission region.\\
	(4): Global covering factor for the [\oiii] emission region .\\
\end{table}

\subsection{Global Covering Factor of the [\oiii] Emission Outflows}
\label{sec:covf}
The global covering factor of the observed [\oiii] emission outflow, i.e., the portion of the full solid angle the [\oiii] emission region covers the central source, can be estimated adopting the same methodology in section 5.2 of \cite{Tian19}. As stated in section \ref{sec:energy}, we assume that within the same quasar, the [\oiii] emission outflow has the same photoionization solutions (i.e., \Uh\ and \Nh) and \ne\ as the BAL outflow. We run the spectral synthesis code Cloudy [version c17.00, \cite{Ferland17}] to predict the expected flux of the [\oiii] \ly 5007 emission emanating from the outflow (F[\oiii]$_\text{mod}$, shown in the second column of table \ref{tab:GlobalCovf}). The observed luminosity of [\oiii] \ly 5007 (L[\oiii]$_\text{obs}$) is shown in the third column of table \ref{tab:GlobalCovf}. Since the [\oiii] emission may be affected by extinction, these L[\oiii]$_\text{obs}$ values are likely to be underestimated. The total illuminating area of the [\oiii] emission region is then:


\begin{equation}\label{eq:size}
 \text{S[O {$\scriptstyle\text{III}$}]} = \text{L[O {$\scriptstyle\text{III}$}]}_\text{obs}/\text{F[O {$\scriptstyle\text{III}$}]}_\text{mod}
\end{equation}

The corresponding global covering factor for the [\oiii] emission region is in the form of: 
\begin{equation}\label{eq:globalC}
\Omega\text{[O {$\scriptstyle\text{III}$}]} = \text{S[O {$\scriptstyle\text{III}$}]}/4\pi R^2 
\end{equation}
where $R$ is the [\oiii] emission outflow's distance from the central source, and we assume the same distance as the BAL outflow from the same object. We show the derived $\Omega$([\oiii]) in the fifth column of table \ref{tab:GlobalCovf}, which is in the range of 1.5\% -- 8.7\%. This $\Omega$([\oiii]) range is consistent with the one derived in \cite{Tian19} ($\sim$4\%), in \cite{Baskin05} (2\% -- 20\%), and in \cite{Dempsey18} ($<$ 50\%). For BAL outflows, this $\Omega$([\oiii]) range is smaller than the global covering factor reported for \civ\ absorption outflows \citep[$\Omega$(\civ) = 20\%, e.g.,][]{Hewett03, Dai08, Dai12, Gibson09, Allen11}, but is similar to that of \siv\ absorption outflows \citep[$\Omega$(\siv) = 8\%]{Borguet13}.

Overall, for all seven quasars in our sample, we find the photoionization solutions and \ne\ derived from the BAL outflows can produce the observed amount of L[\oiii] for the same object with only 1 free parameter [i.e., $\Omega$([\oiii])], while $\Omega$([\oiii]) is similar for all objects within a factor of 6. This strengthen the idea that the gas emitting the [\oiii] has similar \Uh\ and \ne\ to the BAL outflows in the same object.

\section{Summary}
\label{sec:summary}
We present an analysis of 7 BAL/mini-BAL quasar from VLT/X-Shooter observations. These quasars exhibit both absorption and emission outflow features. We present a study of the relationships between these features, and the main results are summarized as follows:

1. For the observed BAL or mini-BAL outflows, we measured physical parameters, such as density, ionization, and distance from the central source. We detect broad and blueshifted [\oiii] emission in 6 out of the 7 objects, which are evidence for emission outflows (section \ref{sec:analysis}).

2. There is a clear trend that the luminosity of the [\oiii] \ly 5007 emission profile decreases with increasing \ne\ of the BAL outflow in the same object (section \ref{sec:neVSLum}). The measured velocity widths from the [\oiii] emission features and \civ\ absorption troughs in the same object are similar (section \ref{sec:compV}), and the mean radial velocity derived from the BAL outflow is moderately larger than the one from the [\oiii] emission outflow (section \ref{sec:compAverV}). Based on these similarities, we have a physical motivation to assume that the gas emitting the [\oiii] has similar \Uh\ and \ne\ to the BAL outflows in the same object. 



3. Comparisons to previous studies show that the objects in our sample exhibit broad [\oiii] emission features similar to the ones in extremely red quasars (selected by extreme infrared-to-optical ratios). Both selection criteria, even though quite different, clearly select objects with high [\oiii] widths and luminosities (section \ref{sec:kine}).

4. By assuming that for the same quasar, the emission outflow has similar physical parameters as the BAL outflow, we constrain the [\oiii] emission region's global covering factor in the range of 1.5\% -- 8.7\%, which is consistent with the ones in the literature (section \ref{sec:covf}).


\section*{Acknowledgements}

X.X., N.A., and T.M. acknowledge support from NSF grant AST 1413319, as well
as NASA STScI grants GO 11686, 12022, 14242, 14054, 14176, and 14777, and NASA ADAP 48020.

Based on observations collected at the European Organisation for
Astronomical Research in the Southern Hemisphere
under ESO programmes: 087.B-0229(A), 090.B-0424(A), 091.B-0324(B), and 092.B-0267(A) (PI: Benn).

CHIANTI is a collaborative project involving George Mason University, the University of Michigan (USA), University of Cambridge (UK) and NASA Goddard Space Flight Center (USA).




\bibliographystyle{mnras}
\bibliography{OIII} 







\appendix
\section{Fits of the Other Objects In The Sample}
\label{sec:APPA}
Here are the fits to the other 6 objects in the sample.
\begin{figure*}

	\includegraphics[width=1\columnwidth,angle=0,clip=true,trim={0.2cm 1.0cm 2.6cm 0.5cm}]{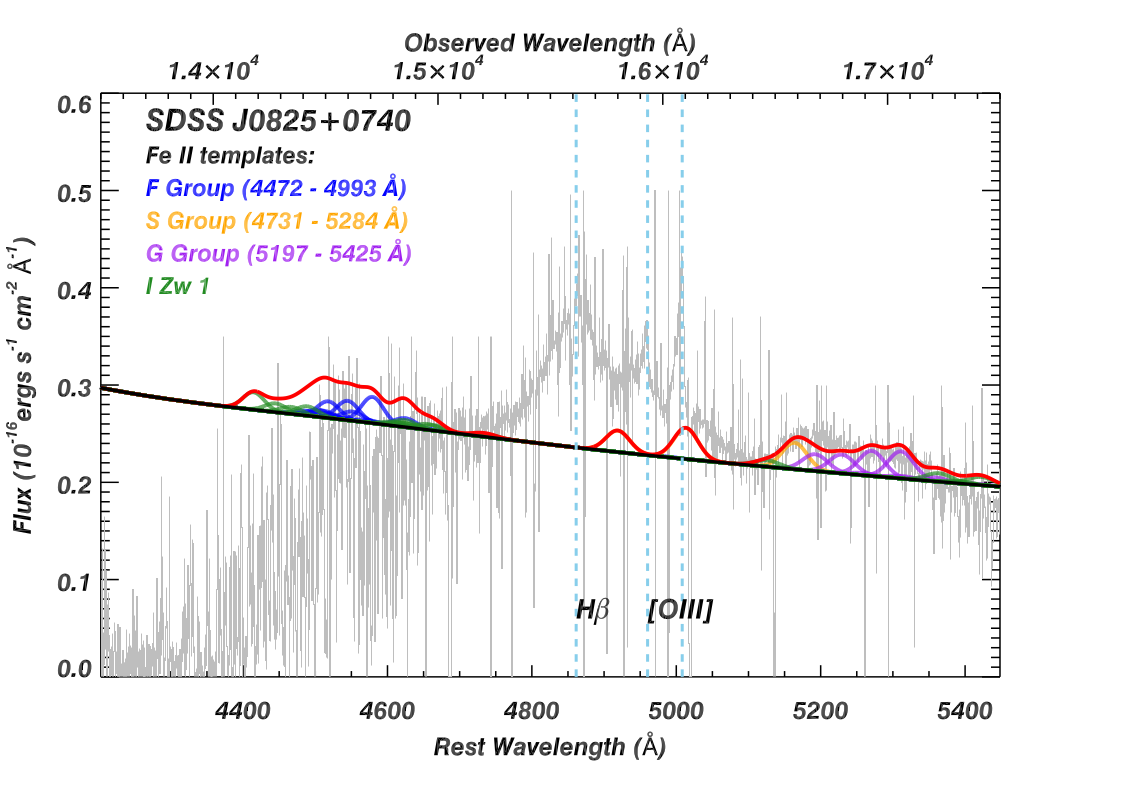}
	\includegraphics[width=1\columnwidth,angle=0,clip=true,trim={0.3cm 1.0cm 2.5cm 0.5cm}]{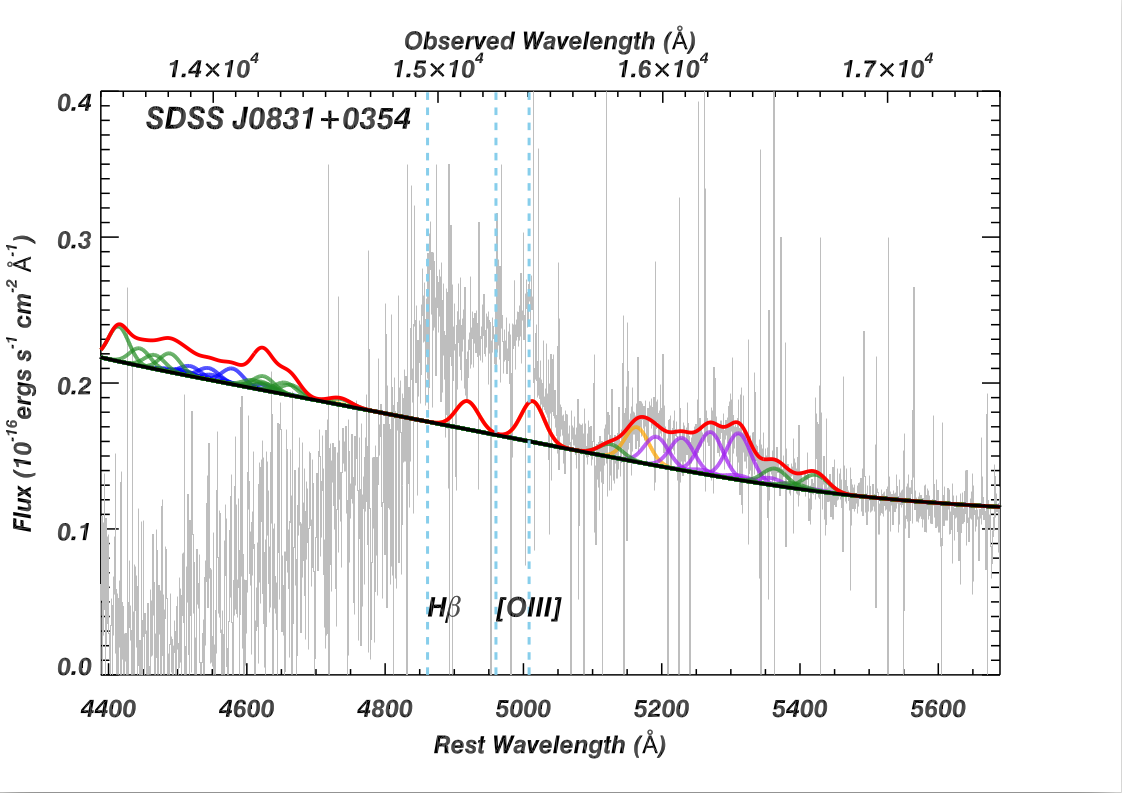}\\

	\includegraphics[width=1\columnwidth,angle=0,clip=true,trim={0.2cm 1.5cm 0.9cm 0.5cm}]{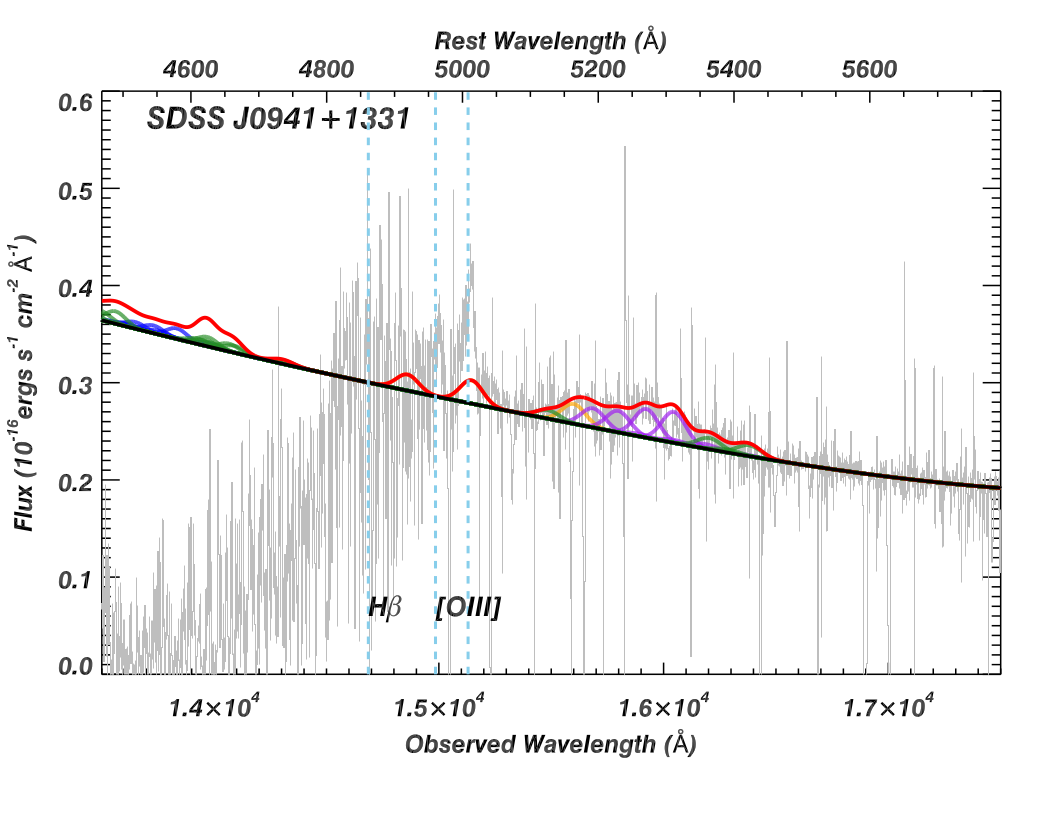}
	\includegraphics[width=1\columnwidth,angle=0,clip=true,trim={0.2cm 1.5cm 0.9cm 0.5cm}]{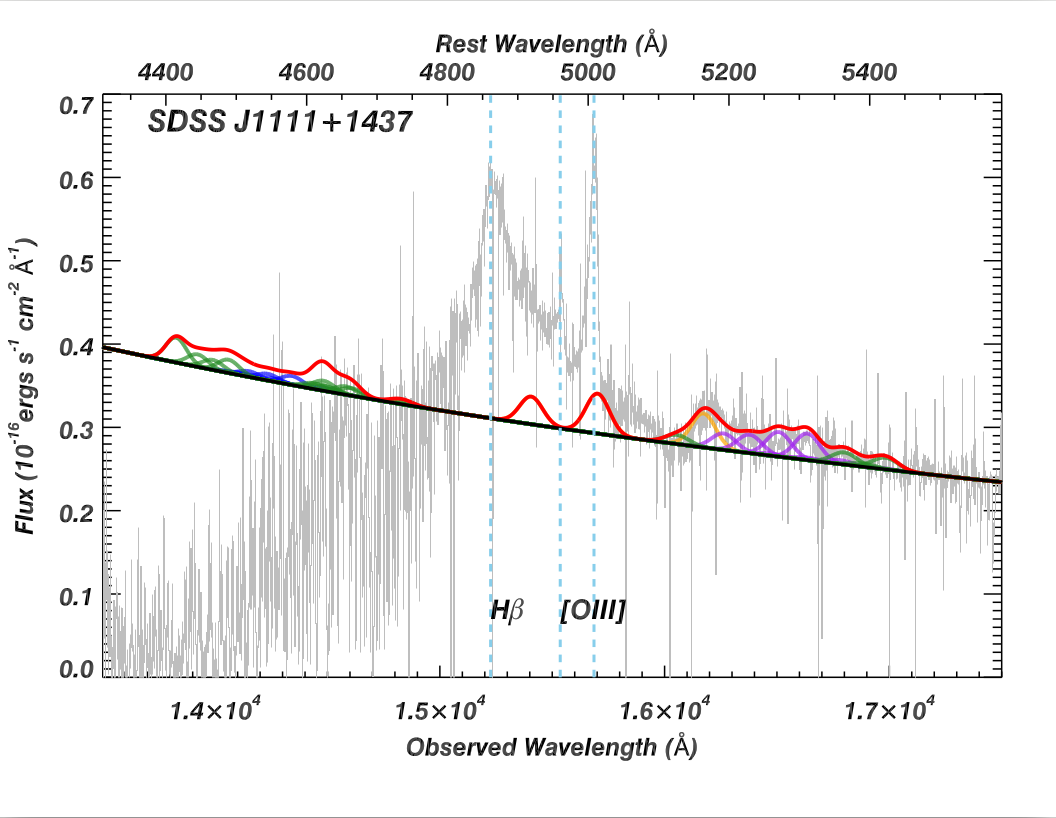}\\

	\includegraphics[width=1\columnwidth,angle=0,clip=true,trim={0.2cm 1.5cm 0.9cm 0.5cm}]{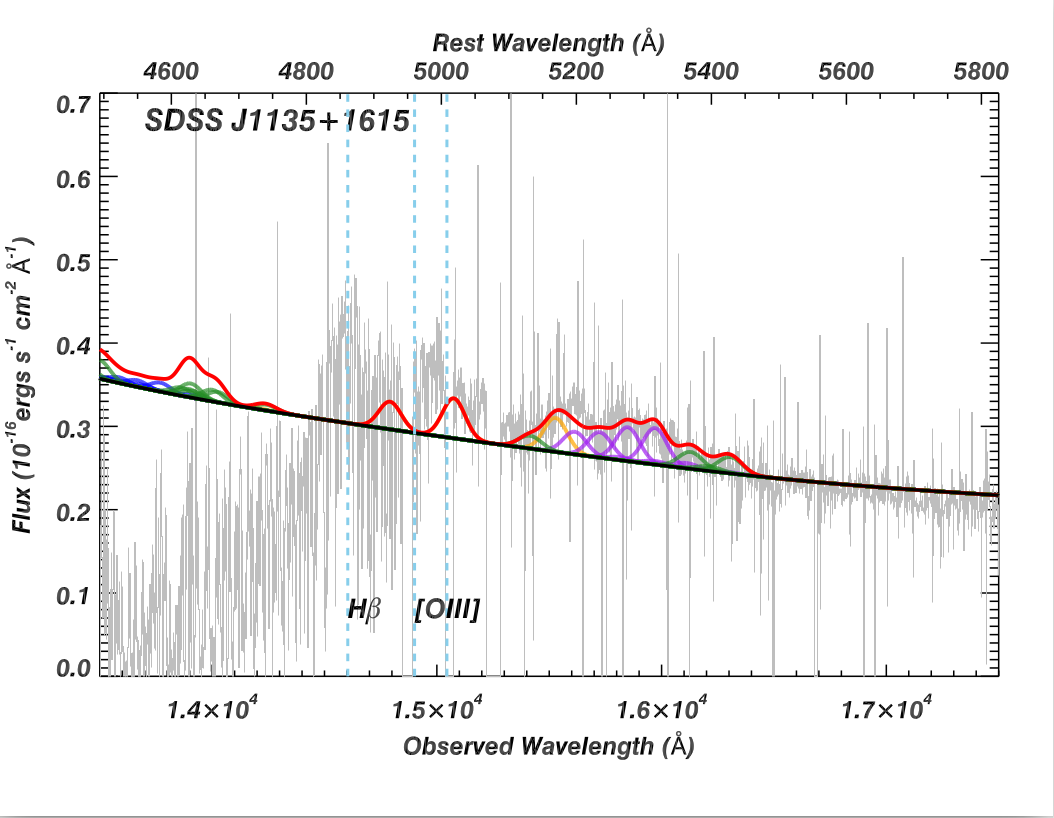}
	\includegraphics[width=1\columnwidth,angle=0,clip=true,trim={0.2cm 1.5cm 0.9cm 0.5cm}]{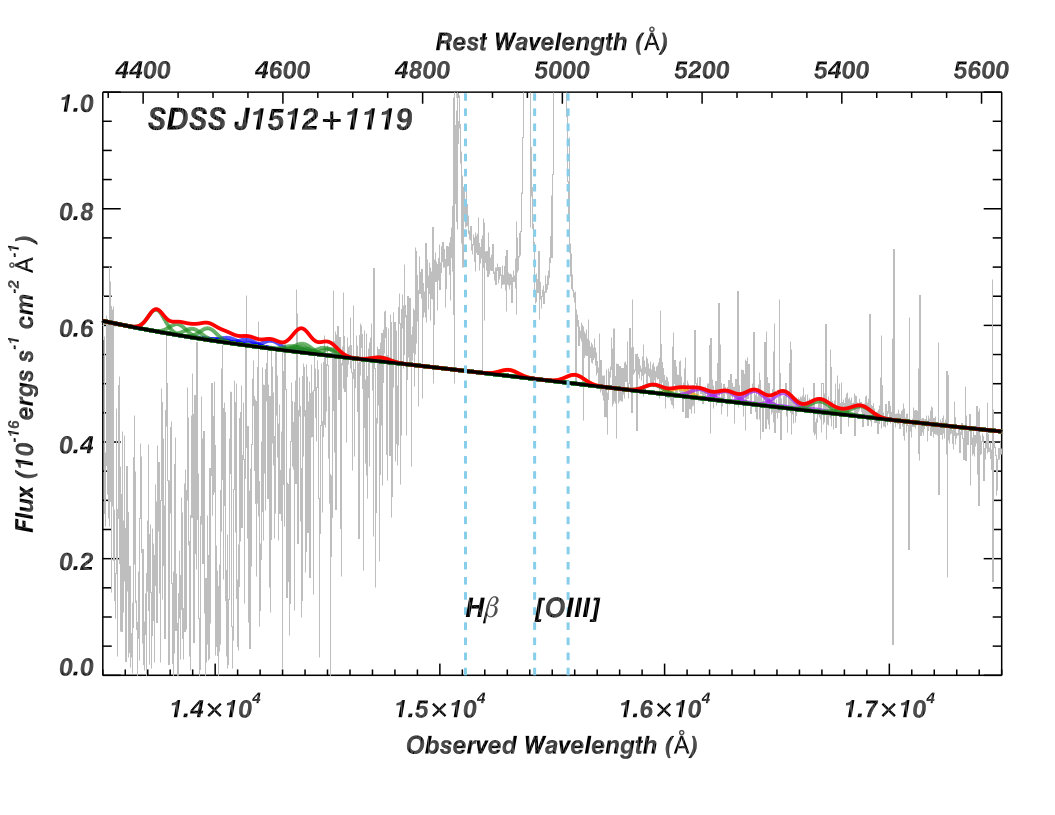}\\
    \label{fig:A1}
\caption{The best fitting \feii\ templates for the other 6 objects in our sample. \feii\ emission profiles are divided into four groups, and are plotted in different colors (see section \ref{sec:FeII} for details). The continuum levels are the same as the ones in figure \ref{fig:OIII_all} and are shown as the black solid lines. The overall \feii\ models created by summing all four groups are shown as solid red lines. The blue dotted lines (left to right) indicate the expected wavelength locations of \hb, [\oiii] \ly 4959, and [\oiii] \ly 5007 in the quasar's rest-frame, respectively. Atmospheric absorption features exist at $\lesssim$ 15,000\angstrom\ in the observed-frame.}
\end{figure*}

\begin{figure*}
	\includegraphics[width=1\columnwidth,angle=0, trim={1cm 1.5cm -0.1cm 0cm}]{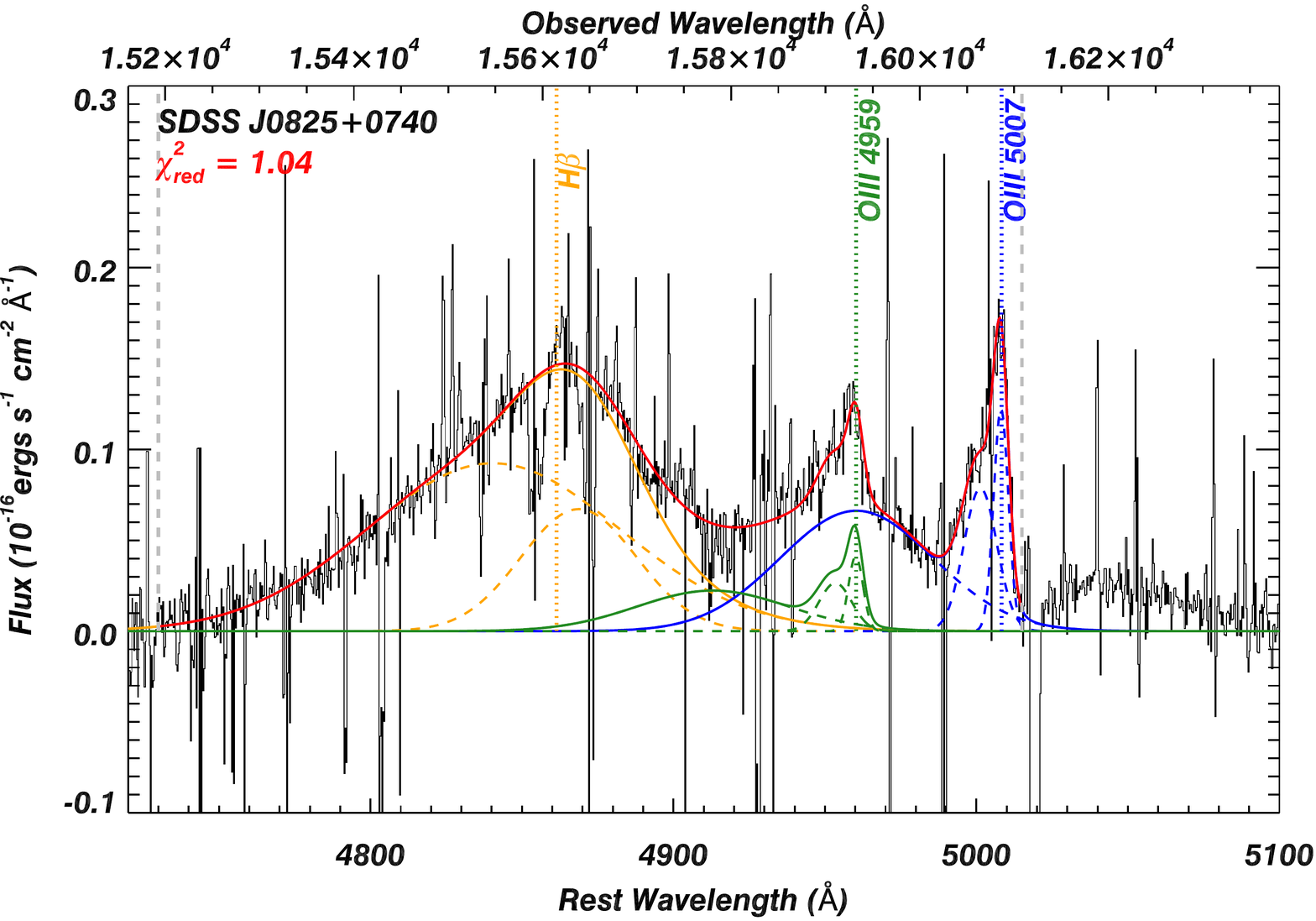}
	\includegraphics[width=1\columnwidth,angle=0, trim={0.0cm 0.9cm 3.0cm 2cm}]{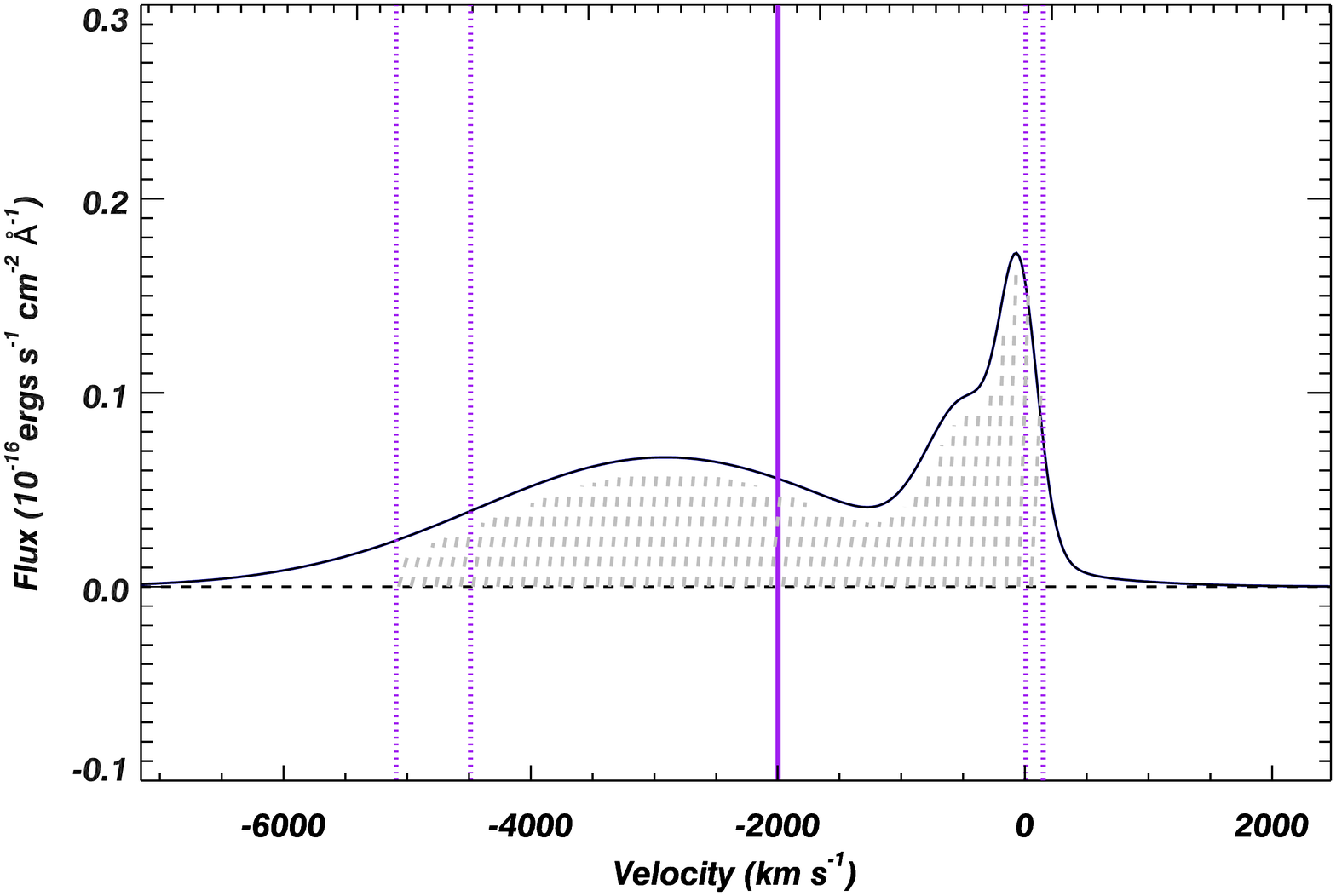}

	\includegraphics[width=1\columnwidth,angle=0,trim={1cm 1.5cm -0.1cm 0cm}]{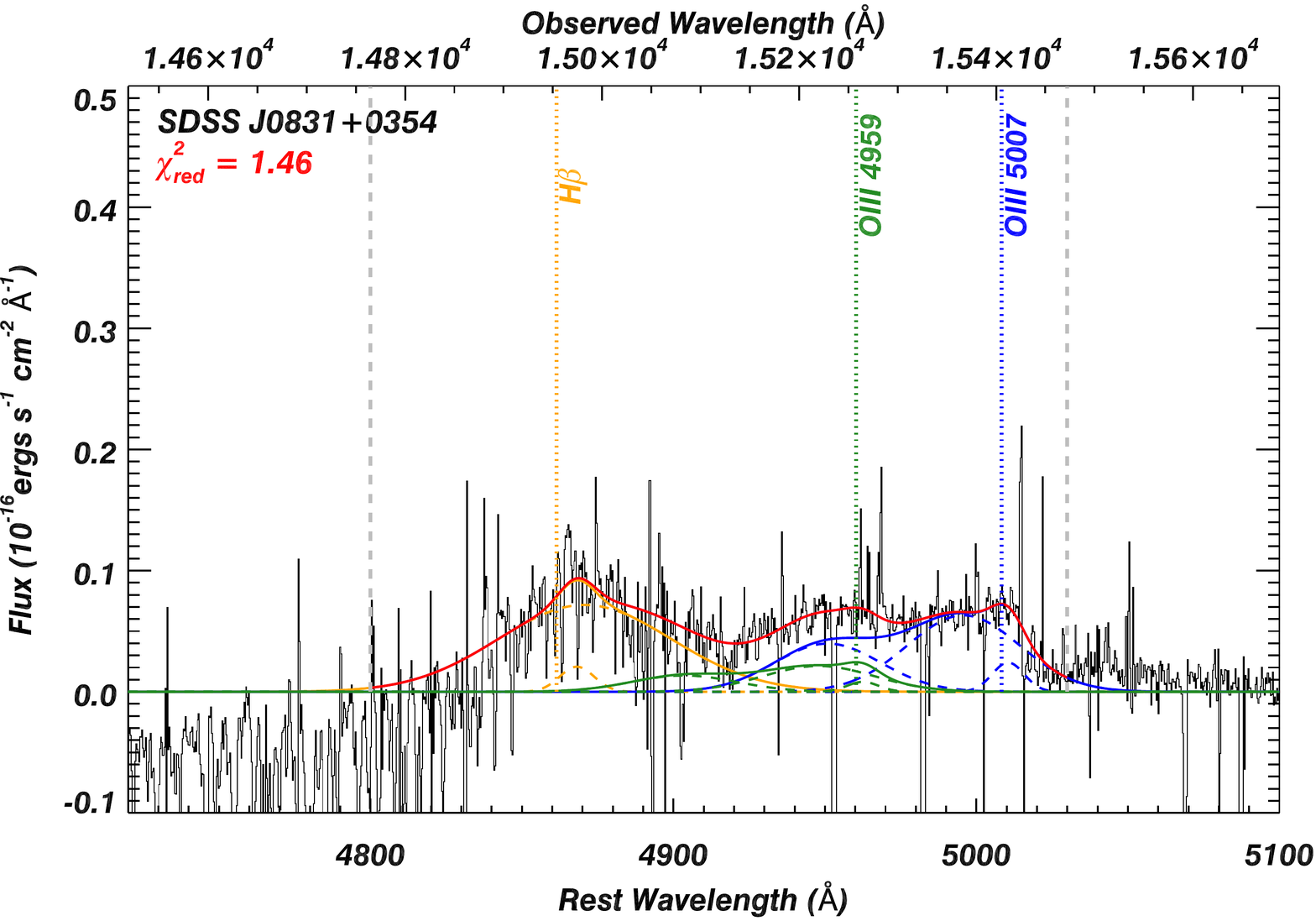}%
	\includegraphics[width=1\columnwidth,angle=0,trim={0.0cm 0.9cm 3.0cm 2cm}]{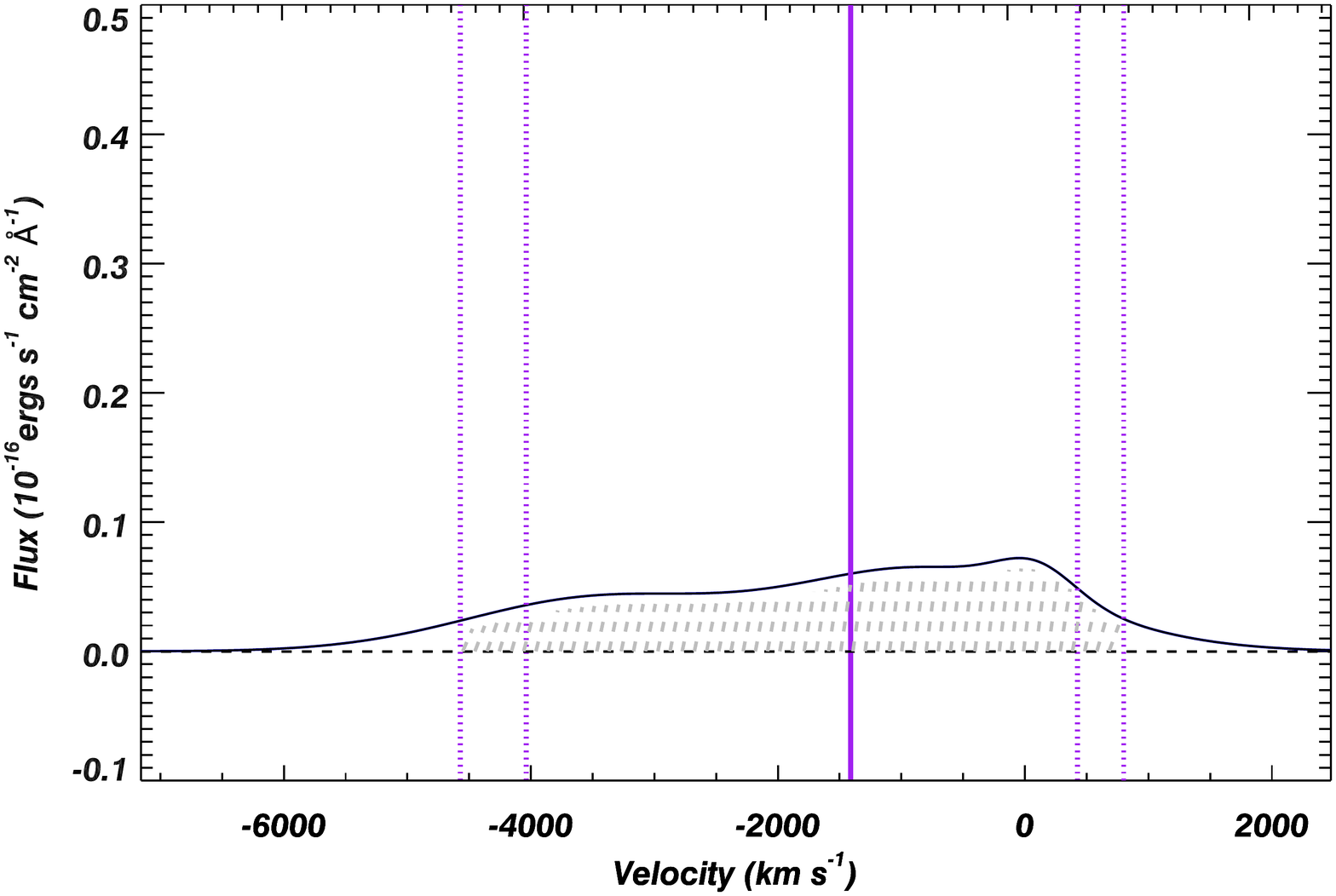}\\

	\includegraphics[width=1\columnwidth,angle=0,trim={1cm 1.5cm -0.1cm 0cm}]{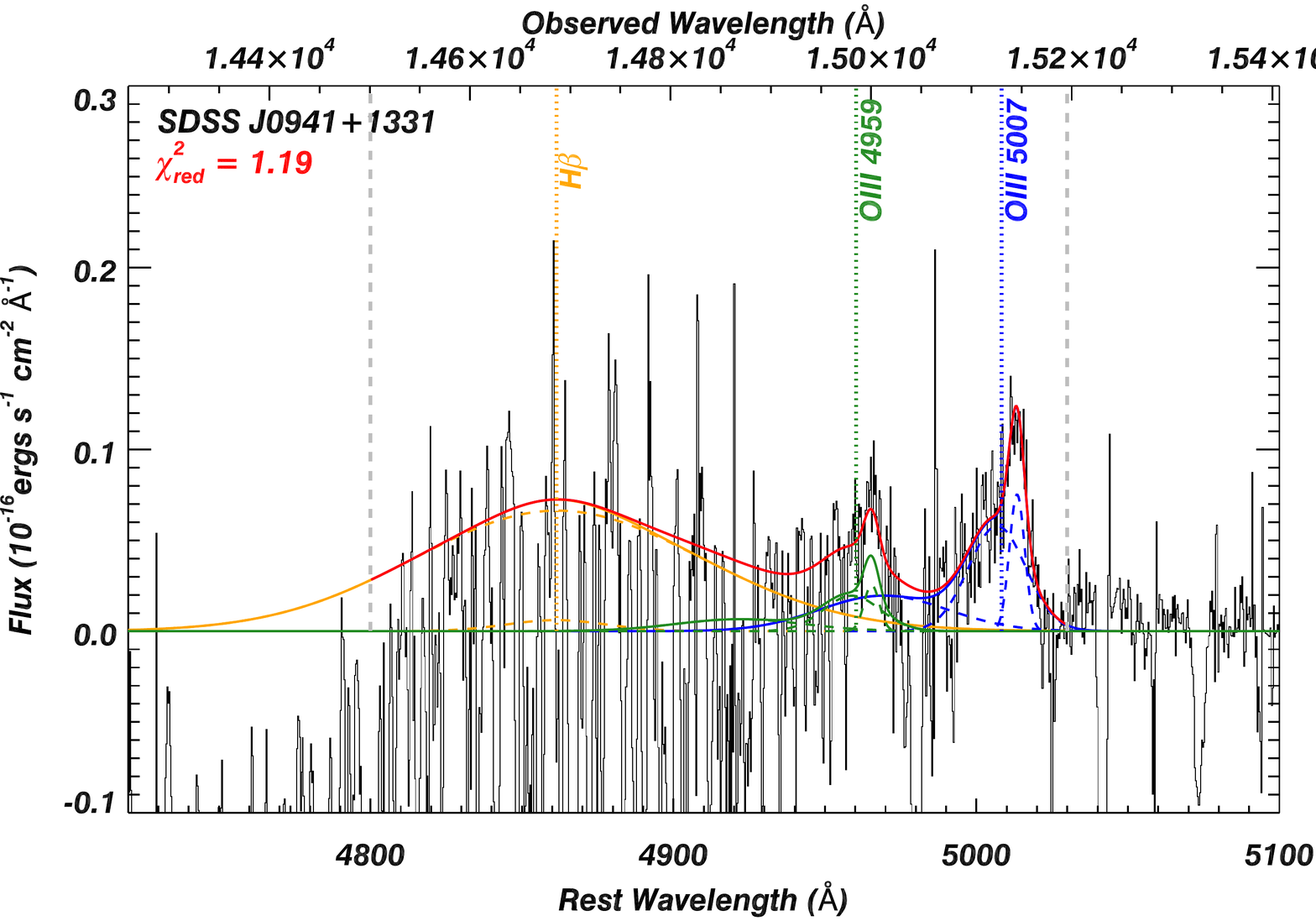}%
	\includegraphics[width=1\columnwidth,angle=0,trim={0.0cm 0.9cm 3.0cm 2cm}]{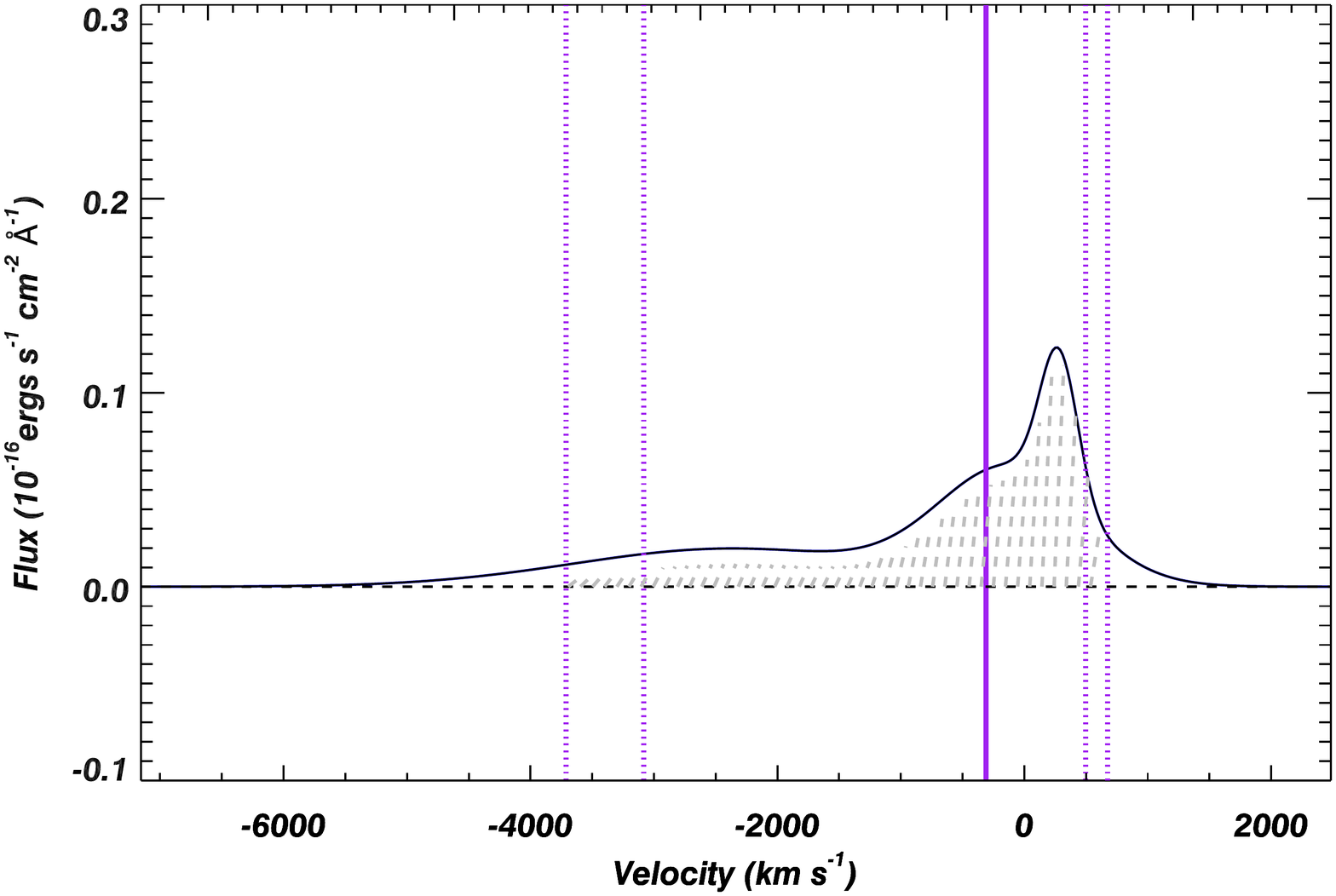}\\

    \label{fig:A2}
\caption{The fits to the \hb\ + [\oiii] emission region \textbf{(Left)} and the extracted [\oiii] \ly 5007 emission profile \textbf{(Right)}. The patterns are the same as figures \ref{fig:J0046all}.}
\end{figure*}


\begin{figure*}

	\includegraphics[width=1\columnwidth,angle=0,trim={1cm 1.5cm -0.1cm 0cm}]{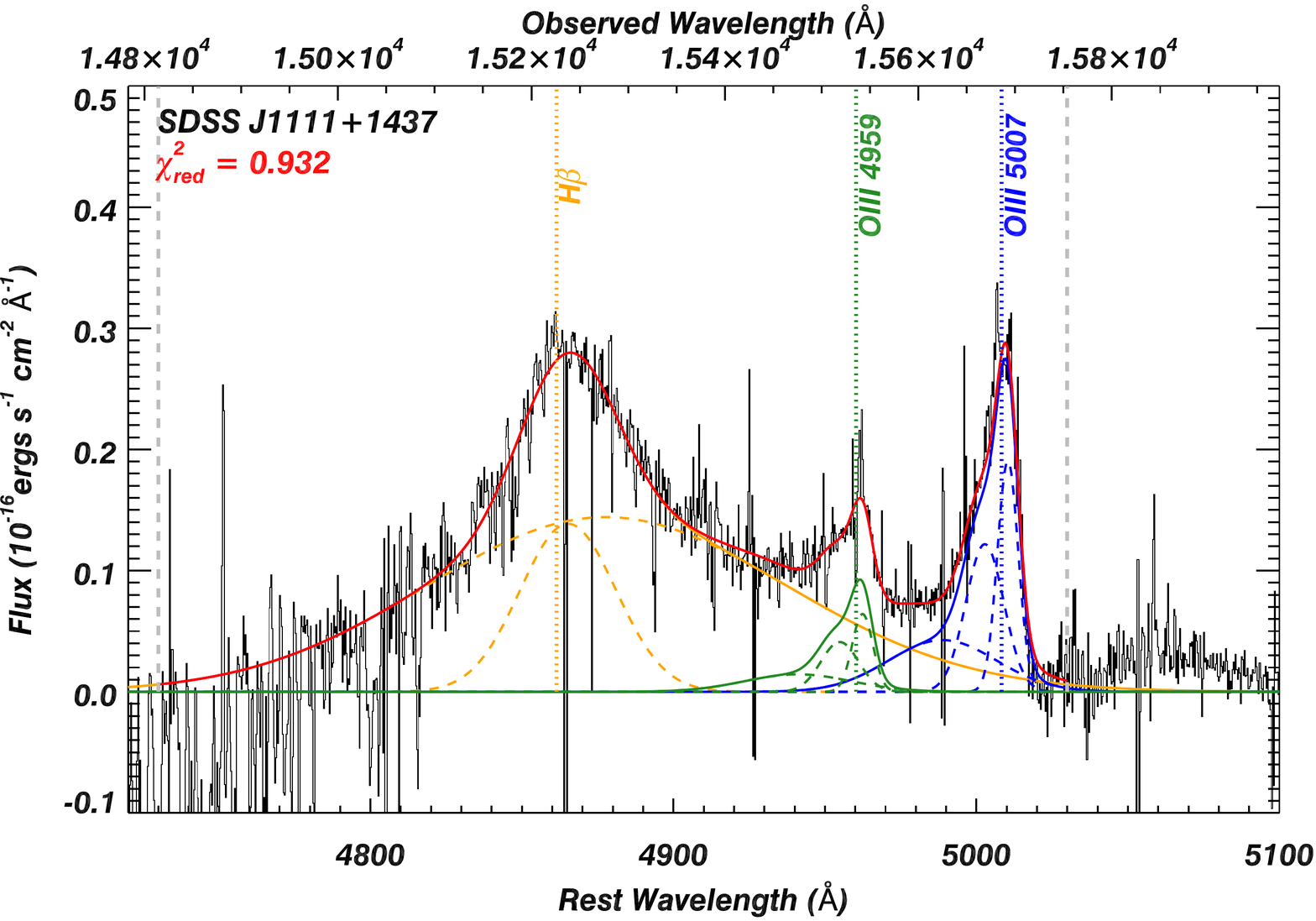}
	\includegraphics[width=1\columnwidth,angle=0,trim={0.0cm 0.9cm 3.0cm 2cm}]{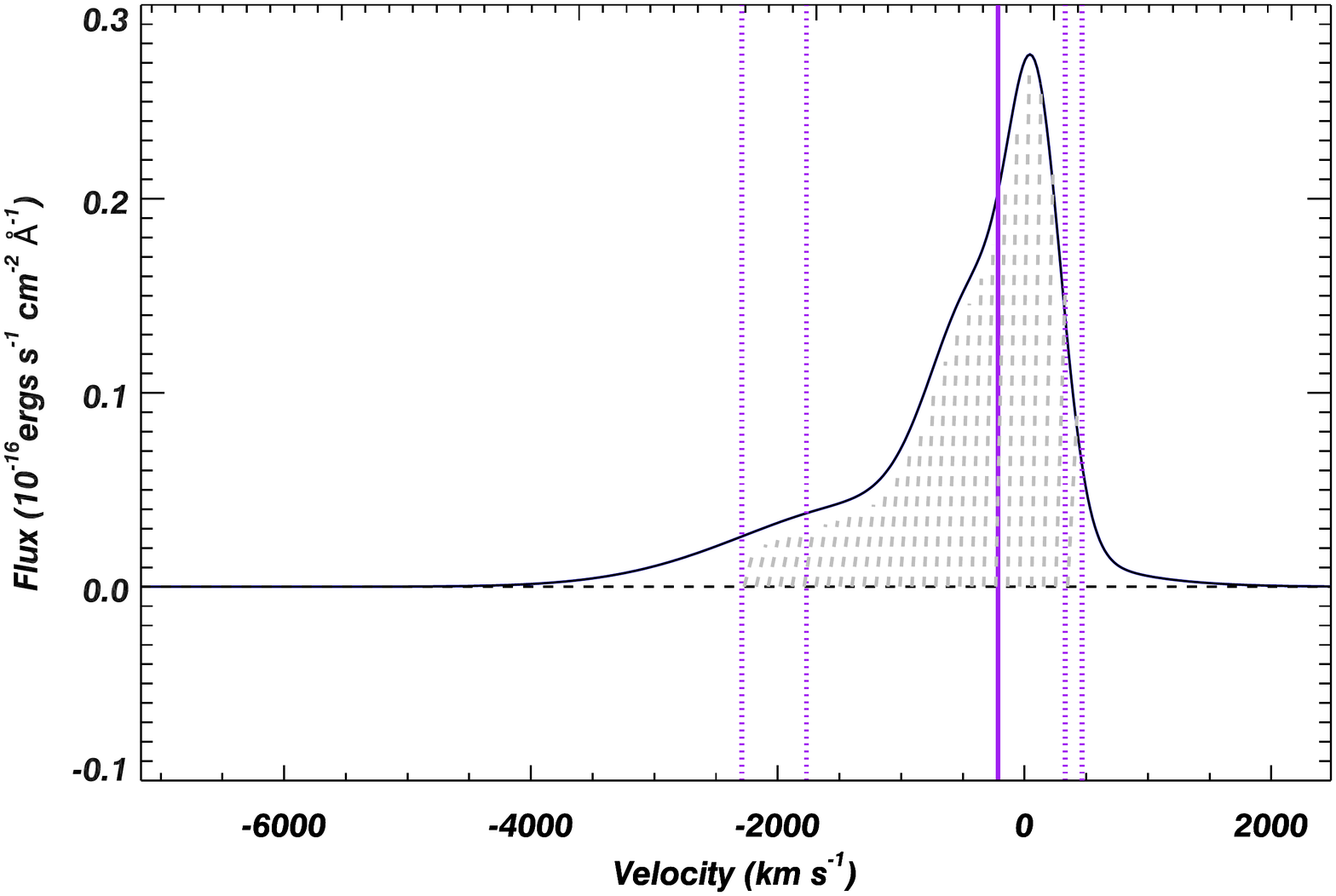}

	\includegraphics[width=1\columnwidth,angle=0,trim={1cm 1.5cm -0.1cm 0cm}]{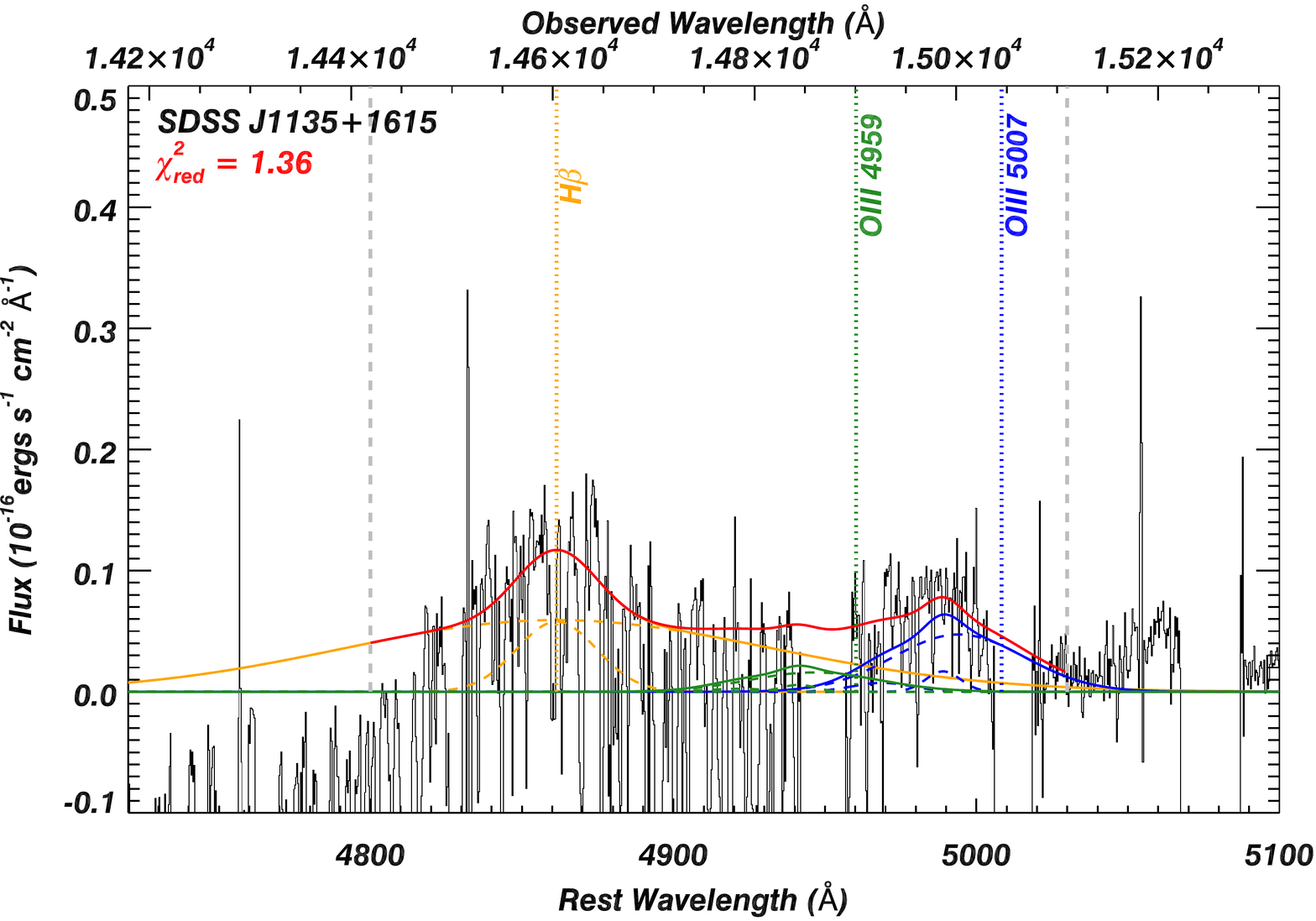}%
	\includegraphics[width=1\columnwidth,angle=0,trim={0.0cm 0.9cm 3.0cm 2cm}]{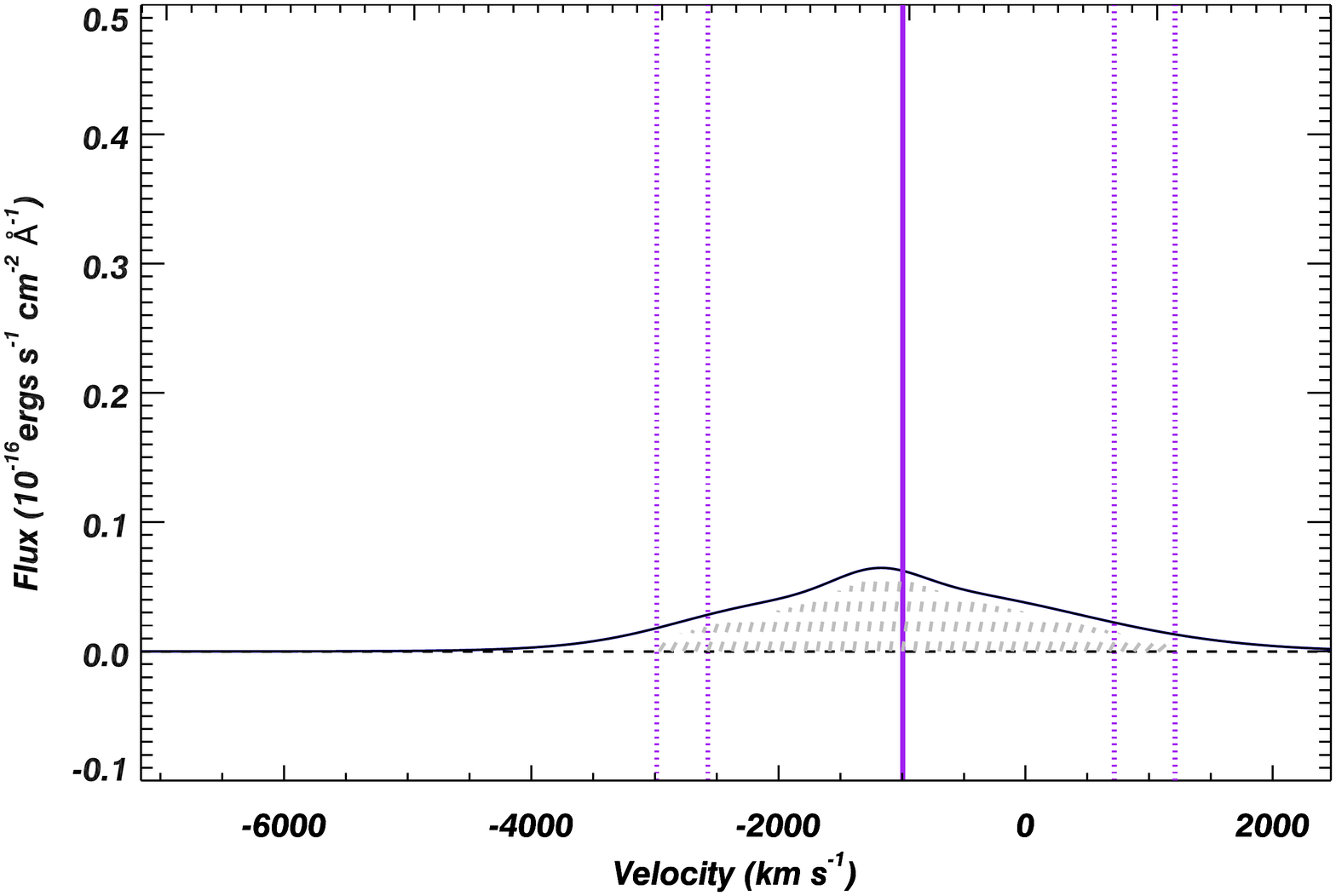}\\

	\includegraphics[width=1\columnwidth,angle=0,trim={1cm 1.5cm -0.1cm 0cm}]{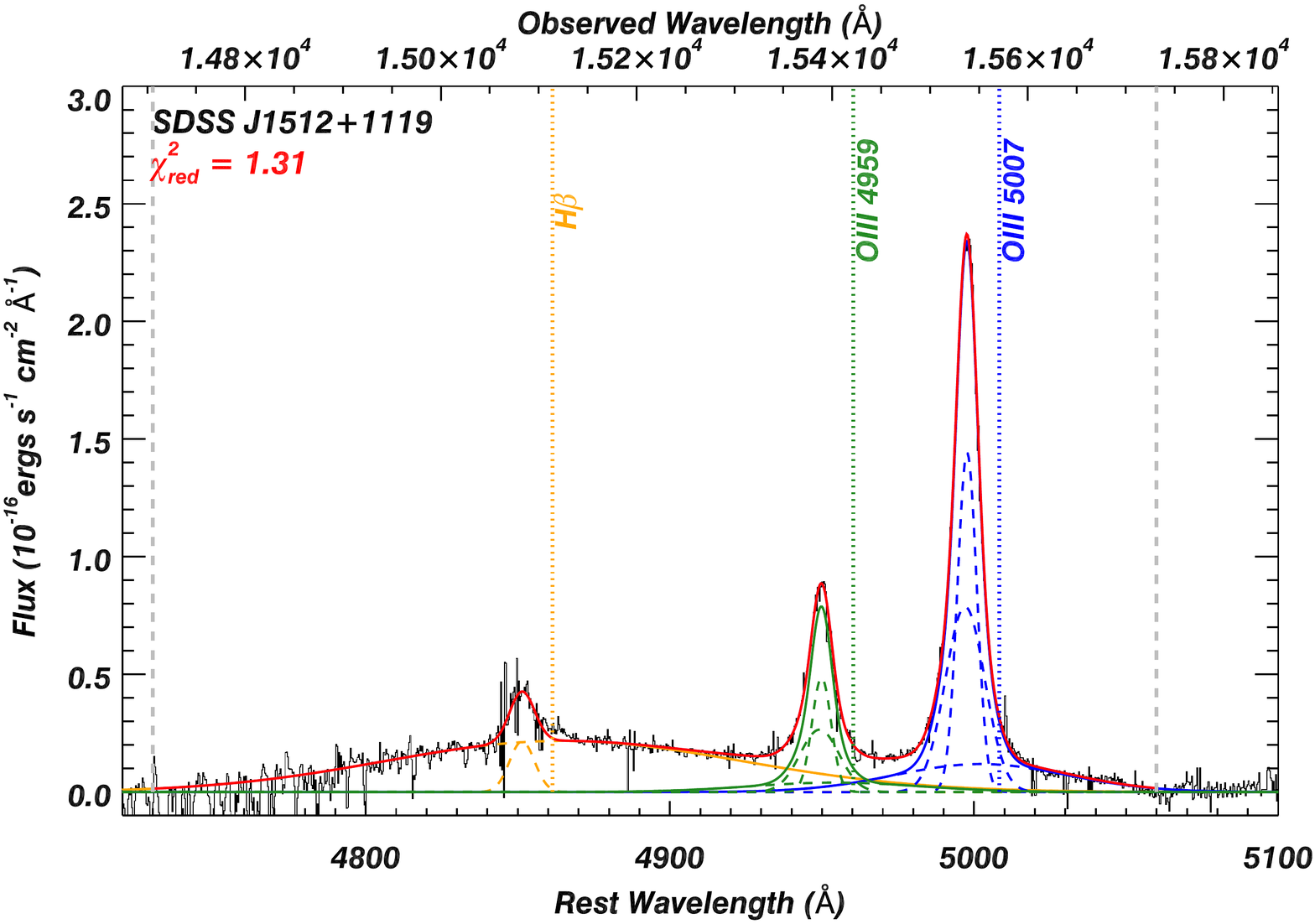}
	\includegraphics[width=1\columnwidth,angle=0,trim={0.0cm 0.9cm 3.0cm 2cm}]{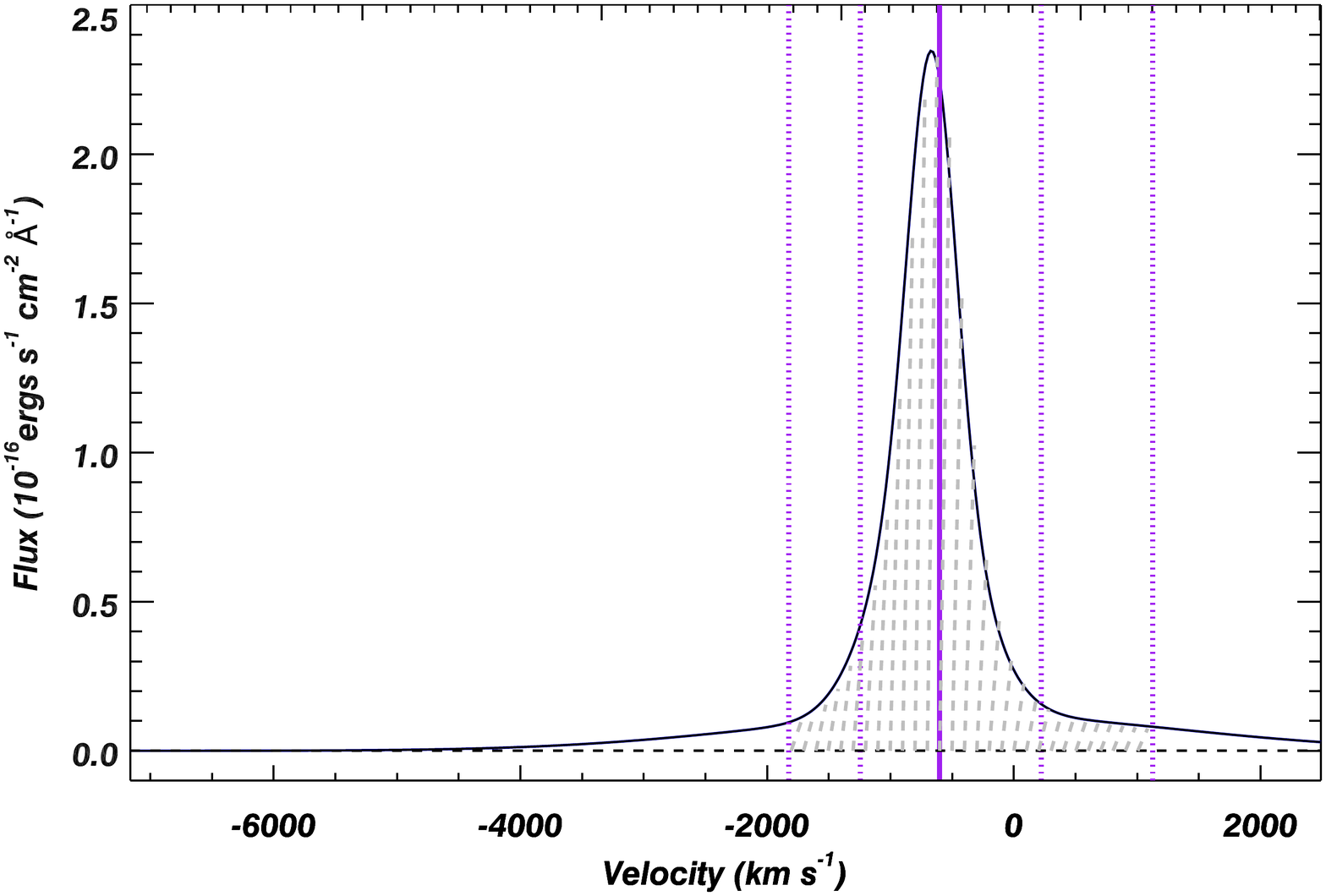}

    \label{fig:A3}
\caption{Same as figure \ref{fig:A2}2. }
\end{figure*}

\bsp	
\label{lastpage}
\end{document}